\documentclass[conference]{IEEEtran}
\IEEEoverridecommandlockouts
\usepackage{subcaption}
\usepackage{cite}
\usepackage{amsmath,amssymb,amsfonts}
\usepackage{algorithmic}
\usepackage{graphicx}
\usepackage{textcomp}
\usepackage{xcolor}
\usepackage{tikz}
\def\BibTeX{{\rm B\kern-.05em{\sc i\kern-.025em b}\kern-.08em
    T\kern-.1667em\lower.7ex\hbox{E}\kern-.125emX}}
\begin{document}

\title{PreSem-Surf: RGB-D Surface Reconstruction with Progressive Semantic Modeling and SG-MLP Pre-Rendering Mechanism\\
}

\author{
    \IEEEauthorblockN{
        Yuyan Ye\textsuperscript{1}, 
        Hang Xu\textsuperscript{1}, 
        Yanghang Huang\textsuperscript{1}, 
        Jiali Huang\textsuperscript{1}, 
        Qian Weng\textsuperscript{1, 2 $*$}
    }
    \IEEEauthorblockA{
        \textsuperscript{1}College of Computer and Data Science, Fuzhou University, Fuzhou 350108, China\\
        \textsuperscript{2}Key Laboratory of Spatial Data Mining and Information Sharing\\ 
        Ministry of Education of the People’s Republic of China, Fuzhou 350108, China
    }
    $^*$ Corresponding author: fzuwq@fzu.edu.cn
}

\maketitle

\begin{abstract}
 We introduce PreSem-Surf, an optimization method based on the Neural Radiance Field (NeRF) framework, in a relatively short time to reconstruct high-quality surfaces from RGB-D sequences of scenes by combining RGB information, depth information, and semantic information.Specifically, we propose a novel sampling structure SG-MLP combined with PR MLP (Preconditioning Multilayer Perceptron) to pre-render voxels, which enables the model to obtain scene-related information earlier, and more effectively distinguish between noise and local details.PreSem-Surf, compared with existing models, achieves a better balance between smoothness and accuracy of reconstruction. In addition, precision-progressive semantic modeling is introduced to extract semantic information with progressive levels of accuracy, enabling the model to learn scene information while minimizing training time.The model is trained and evaluated by means of seven scenes and six metrics from synthetic datasets, allowing for comprehensive benchmarking. On average across all scenes, our model achieved the best performance in terms of C-L1, F-score, and IoU, with its performance in NC, Acc, and Comp slightly behind the best models.
\end{abstract}


\section{Introduction}
In recent years, the demand for 3D scene understanding and virtual environment visualization has been continuously increasing. Traditional 3D reconstruction methods, such as Multi-View Stereo and Phase Shifting Algorithms \cite{Sinha2007,Zuo2018}, have achieved automated 3D data acquisition and processing. However, these methods often suffer from limited accuracy, sensitivity to the quality of input information, and high resource consumption. The implicit neural representation (NeRF) \cite{Mildenhall2021} proposed by Mildenhall et al. is of great significance. It uses a multilayer perceptron (MLP) to predict the volumetric density and color of spatial points, optimizing the scene representation. This method has achieved significant results in high-quality view rendering of complex scenes, with lower storage costs and a certain ability to model unseen objects.
\begin{figure}[t]
    \centerline{\includegraphics[width=0.9\columnwidth]{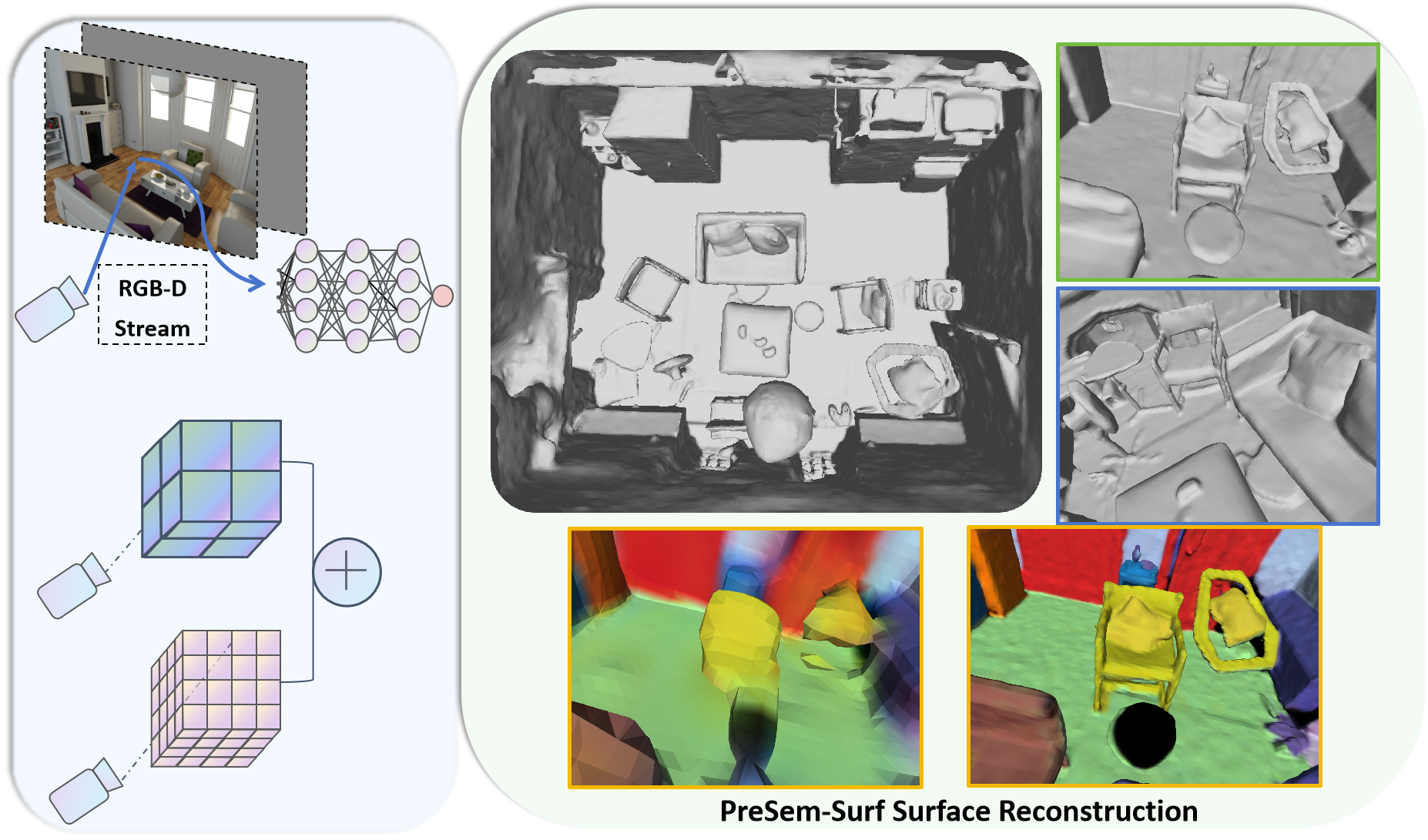}}
    \caption{Building upon the neural radiance fields (NeRF), we propose PreSem-Surf, which integrates voxel rendering mechanisms with multimodal information such as scene semantics.}
    \label{fig:presemsurf}
\end{figure} 
However, NeRF still faces many challenges in practical applications. On the one hand, its implicit representation requires high-quality data. Low-quality input can easily lead to blurred modeling results, artifacts, or even fragmented models. To address this, incorporating semantic information can enhance its ability to infer missing information, thus further improving the model's generalization capability \cite{Zhu2024,Li2023,Duncan2021}. Alternatively, using light models for preprocessing before rendering can help provide reference information for the main model during training \cite{Tu2022,Zhu2022}. This not only reduces training time but also helps avoid local optima, improving reconstruction quality. However, both semantic information and preprocessing inevitably consume
additional resources, and their impact on reconstruction quality is very complex. Inappropriate introduction of semantic information may negatively affect certain metrics of the reconstruction results. Therefore, developing a
NeRF model that can effectively utilize semantic information and preprocessing is both necessary and challenging.

On the other hand, NeRF still has high resource consumption. Although its storage requirements are reduced compared to traditional methods, the network parameters and intermediate features still occupy a significant amount of memory \cite{Mildenhall2021}. Additionally, its reliance on MLP as the main structure makes the training time far from satisfactory. To address this, sparse networks such as octrees \cite{Yang2022}, multi-resolution hash tables \cite{Wang2023}, or mapping networks \cite{Duncan2021}  can be used to simplify scene information and model structure. Signed Distance Function(SDF) can be introduced for small-scale scenes \cite{Azinovic2022,Wang2022go,Johari2023}, or the MLP architecture can be reduced \cite{Wang2022go}. However, discarding some information will inevitably affect the reconstruction quality, regardless of the method used. Therefore, finding a way to balance resource consumption and reconstruction quality remains a key issue in this field.

Based on the aforementioned limitations, this paper proposes the PreSem-surf method. To address the issues, this paper designs a decoder structure, SG-MLP, by pre-rendering voxels and integrating scene semantic information to perform layered pre-rendering before the actual rendering. Combined with a progressive semantic modeling strategy, the PreSem-surf method effectively utilizes various aspects of scene information to enhance sampling efficiency, rendering quality, accuracy, and the completeness and precision of scene reconstruction.
We conducted experiments on the Synthetic Dataset to verify the superiority of our method. In summary, the contributions of this paper are as follows:
\begin{itemize}
    \item A novel sampling-guided multilayer perceptron is proposed, which pre-renders the scene based on the original NeRF architecture. Through a unique layered pre-rendering mechanism, it provides critical guidance for subsequent formal rendering, enabling the model to efficiently and accurately reconstruct the scene. This significantly improves the quality and efficiency of scene reconstruction, overcoming the deficiencies of traditional methods in sampling and rendering.
    \item A progressive semantic modeling strategy is designed, following the logical sequence of "perception-semantic-segmentation-modeling," and gradually refines the modeling process according to scene semantics. This strategy significantly enhances the completeness and smoothness of scene reconstruction, making the results more realistic.
\end{itemize}

This paper conducted comparative experiments and ablation analysis on the Synthetic Dataset public dataset. By comparing with various advanced methods, it fully demonstrated the effectiveness of each module of the proposed method and clearly explained the operating mechanisms of each module in the entire scene reconstruction process.
\section{Related Work}
\textbf{3D Reconstruction:} In the field of 3D reconstruction, NeRF and 3D Gaussian Splatting (3DGS) are two important techniques each with their own characteristics and application scenarios. NeRF represents scenes as 5D neural radiance fields, successfully overcoming the limitations of previous methods and achieving high-quality view synthesis \cite{Mildenhall2021}. 3DGS is a 3D reconstruction technique of real-time radiance field rendering. Its core idea is to use 3D Gaussian distributions as volume representations to model scenes and achieve efficient rendering \cite{Kerbl2023}.
Compared to NeRF, 3DGS has faster rendering speeds but requires substantial storage resources and is sensitive to input quality. Lee Chan et al. significantly improved memory and storage efficiency through innovative volume masking strategies and compact attribute representations \cite{Lee2024}. Mip-Splatting addressed aliasing issues in 3D Gaussian rendering by introducing 3D smoothing filters and 2D Mip filters, demonstrating excellent performance across different scales \cite{Yu2024}. In addition to the above points, the MLP-based structure of NeRF offers better extensibility and has had a profound impact on various subfields of 3D computer vision, such as novel view synthesis \cite{Martin-Brualla2021,Mildenhall2022,Mildenhall2021,Verbin2022}, surface reconstruction \cite{Park2019,Wang2022,Wang2021neus}, dynamic scene representation \cite{Park2021,Pumarola2021}, and camera and pose estimation \cite{Lin2021,Wang2021,Xia2022,Yen-Chen2021}.
\par

\textbf{Signed Distance Function(SDF):} In the field of 3D reconstruction, the Signed Distance Function (SDF) has emerged as a powerful representation method, widely applied in several state-of-the-art techniques. DeepSDF utilizes neural networks to predict SDF values from 3D points to surfaces, leveraging latent space encoding to achieve efficient reconstruction of complete shapes from partial observations \cite{Park2019}. SDF-SRN focuses on 3D shape reconstruction from single RGB images, employing differentiable rendering to optimize SDFs and recover more accurate 3D shapes and topologies from images \cite{Yu2024gsdf}. GO-Surf enhances SDF applications by optimizing hierarchical feature grids and SDF values for fast, high-fidelity surface reconstruction from RGB-D sequences, and a novel SDF gradient regularization term is introduced to aid in hole filling and detail preservation \cite{Wang2022go}. GSDF integrates 3D Gaussian Splatting (3DGS) with neural SDFs in a dual-branch architecture, significantly improving rendering quality and geometric reconstruction accuracy through joint supervision. Together, these methods have advanced the use of SDF in 3D reconstruction, offering new possibilities for efficient and high-quality reconstruction of complex scenes \cite{Yu2024gsdf}.
%here
\begin{figure*}[t]
    \centering
    \includegraphics[width=0.85\linewidth]{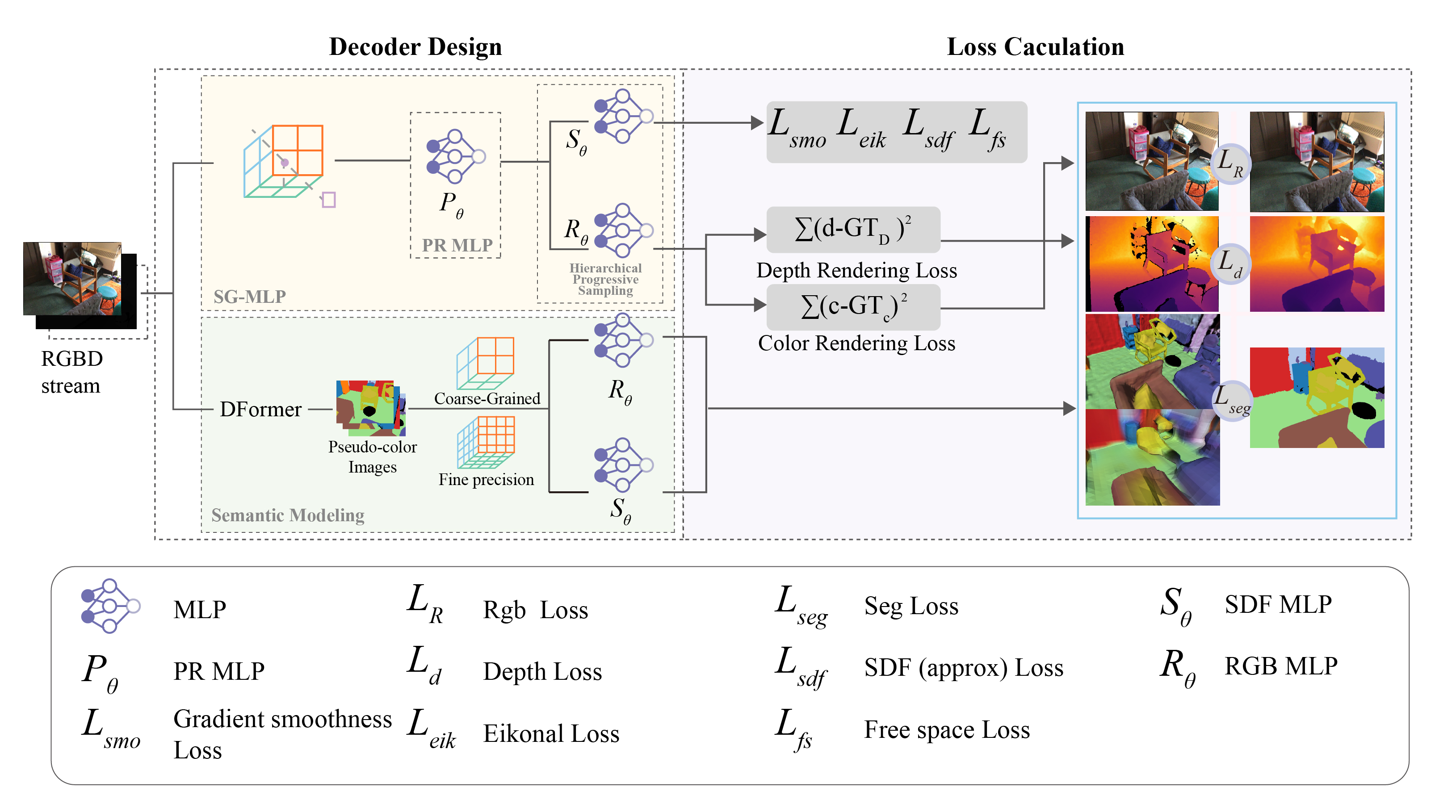}
    \caption{Overview of PreSem-Surf. SG-MLP first uses PR MLP for coarse volumetric density estimation without color rendering, and then gradually improves the accuracy. Then, two MLPs process scene RGB and depth information, and Dformer generates pseudo-color images for coarse-to- fine scene rendering. Finally, loss functions adjust the model's reconstruction from different perspectives.}
    \label{fig:struct}
\end{figure*}

\par
\textbf{Semantic Information:} In 3D reconstruction, semantic information plays a significant role. It assists reconstruction algorithms in understanding scenes,
improving the accuracy of geometric reconstruction, enhancing the robustness of algorithms under noisy and incomplete data conditions, and assists in optimize the reconstruction process to achieve real-time or near-real-time 3D reconstruction \cite{Duncan2021,Zhu2024}. In MSeg3D \cite{Li2023}, semantic information enables the model to more accurately fuse LiDAR and camera features, thereby improving segmentation accuracy. 
SNI-SLAM \cite{Zhu2024} improves the representation of features by combining semantic information with appearance and geometric features through a mechanism of cross attention, allowing the system to remain robust even when a single attribute is defective. In Kimera \cite{Rosinol2020}, semantic information provides a higher level of abstraction and more precise environmental modeling, enabling robots to recognize and understand objects and structures in the scene.

The aforementioned models optimize and enhance NeRF from various perspectives, but currently lack a model that can effectively integrate these methods to further leverage their strengths. Based on these observations, we propose a new voxel rendering mechanism combined with SDF to extract RGB information, geometric information, and semantic information from the scene. This approach improves training efficiency and efficiently utilizes various types of scene information, further enhancing reconstruction quality while minimizing resource consumption.

\section{Method}
Our method is outlined in Fig. \ref{fig:struct}. SG-MLP utilizes PR MLP for uniform pre-sampling, followed by RGB MLP and SDF MLP in SG-MLP. These two components adopt a layered progressive sampling strategy. PFPSMS quickly constructs the overall framework through coarse-grained rendering, capturing the main structures, and then progressively refines the details through fine-grained rendering.
\subsection{Sampling-Guided Structure SG-MLP}
In the light rendering process, we identify two stages. The primary stage aims to quickly acquire an approximate representation of the scene's voxels, enabling the rapid construction of the scene's basic structure. Specifically, a set of sampling points ${x_i}_{i=1}^N$ is used, and SG-MLP employs a simplified MLP network along with a uniform sampling strategy to estimate the initial volume density at each sampling point $x_i$. The output of this process is $\sigma_i$, as follows:
\begin{equation}
\sigma_i = \text{MLP}_\theta \big(\gamma(x_i)\big)
\end{equation}
where $\text{MLP}_\theta$ represents the PR MLP, a simplified MLP network, and $\gamma(x_i)$ is the high-frequency encoding of the sampling point $x_i$, which enhances spatial information through the frequency encoding function $\gamma$ used in NeRF.

In the initial stage, SG-MLP is used to perform a rough volume density estimation, and quickly construct the basic framework of the scene. Although the initial sampling strategy is efficient, it cannot capture the scene's details and features, thus making it inadequate for high-precision rendering. This is where the hierarchical progressive sampling strategy comes into play.

The core idea of the hierarchical progressive sampling strategy is that each subsequent sampling layer refines the distribution of the volume density $\sigma_i$ based on the estimates from the previous layer. To improve sampling resolution, after the initial volume density estimation, the SG-MLP adjusts the sampling strategy for the next layer. For example, at the $(k+1)$-th layer, assuming there are $N$ sampling points, SG-MLP computes a dynamic threshold $\tau_{k+1}$ based on the volume density distribution of the $k$-th layer’s sampling points. This threshold is then used to select key regions for sampling. The calculation formula is as follows:
\begin{equation}
\tau_{k+1} = \lambda \cdot \frac{1}{N} \sum_{j=1}^{N} \sigma_k(x_j) + (1-\lambda) \cdot \max_{1 \leq j \leq N} \sigma_k(x_j)
\end{equation}
where $\lambda$ is a weight parameter that controls whether the threshold is more influenced by the mean or the maximum value, and $\sigma_k(x_j)$ represents the volume density at the $x_j$-th sampling point in the $k$-th layer.
\par
After calculating the dynamic threshold $\tau_{k+1}$, we filter the sampling point set, keeping only the points where $\sigma(x_d) > \tau_{k+1}$. Next, within the retained sampling points, importance sampling is performed based on the volume density, prioritizing the generation of sampling points in regions with higher density. This ensures that important regions are computed with greater precision during the rendering process. The probability density function for generating the next layer of sampling points $p(x_d)$ is calculated as follows:
\begin{equation}
p_{k+1}(x_d) = \frac{\sigma_k(x_d)}{\sum_{j \mid \sigma_k(x_j) > \tau_{k+1}} \sigma_k(x_j)}
\end{equation}
where $\sum_{j \mid \sigma_k(x_j) > \tau_{k+1}} \sigma_k(x_j)$ represents the sum of the volume densities of all the retained sampling points.
\par
After completing the initial coarse sampling, the subsequent rendering process fully utilizes these preliminary estimates as guidance. The final color and density estimates adopt the same standard formulas as GO-Surf. Through experiments, it has been demonstrated that this mechanism, guided by hierarchical pre-rendering, achieves efficient scene reconstruction and high-precision rendering through progressively optimized sampling strategies.
\subsection{Progressive semantic modeling strategy}
Raditional 3D reconstruction methods often use a uniform modeling strategy, which fails to flexibly address the complexity of different levels and details within a scene. We have found that a progressive modeling approach is more effective for semantic representation of the environment. When observing a complex scene, we typically begin by understanding its overall layout, identifying major objects and structures, and forming a rough understanding. Gradually, our attention shifts to finer local features, enriching our perception of the scene. Inspired by this process, the progressive semantic modeling strategy adopts a step-by-step approach of "perception-semantic-segmentation-modeling." It first captures the scene's global structure and then refines local details, improving efficiency and accuracy for precise, high-quality reconstruction.
\begin{figure}[!t]
    \centering
    \begin{minipage}{0.8\columnwidth} % 整体宽度接近单栏宽度，留一点边距
        \centering
        \begin{subfigure}[t]{0.48\textwidth} % 每个子图占约48%宽度，留一点间隙
            \centering
            \includegraphics[width=\textwidth]{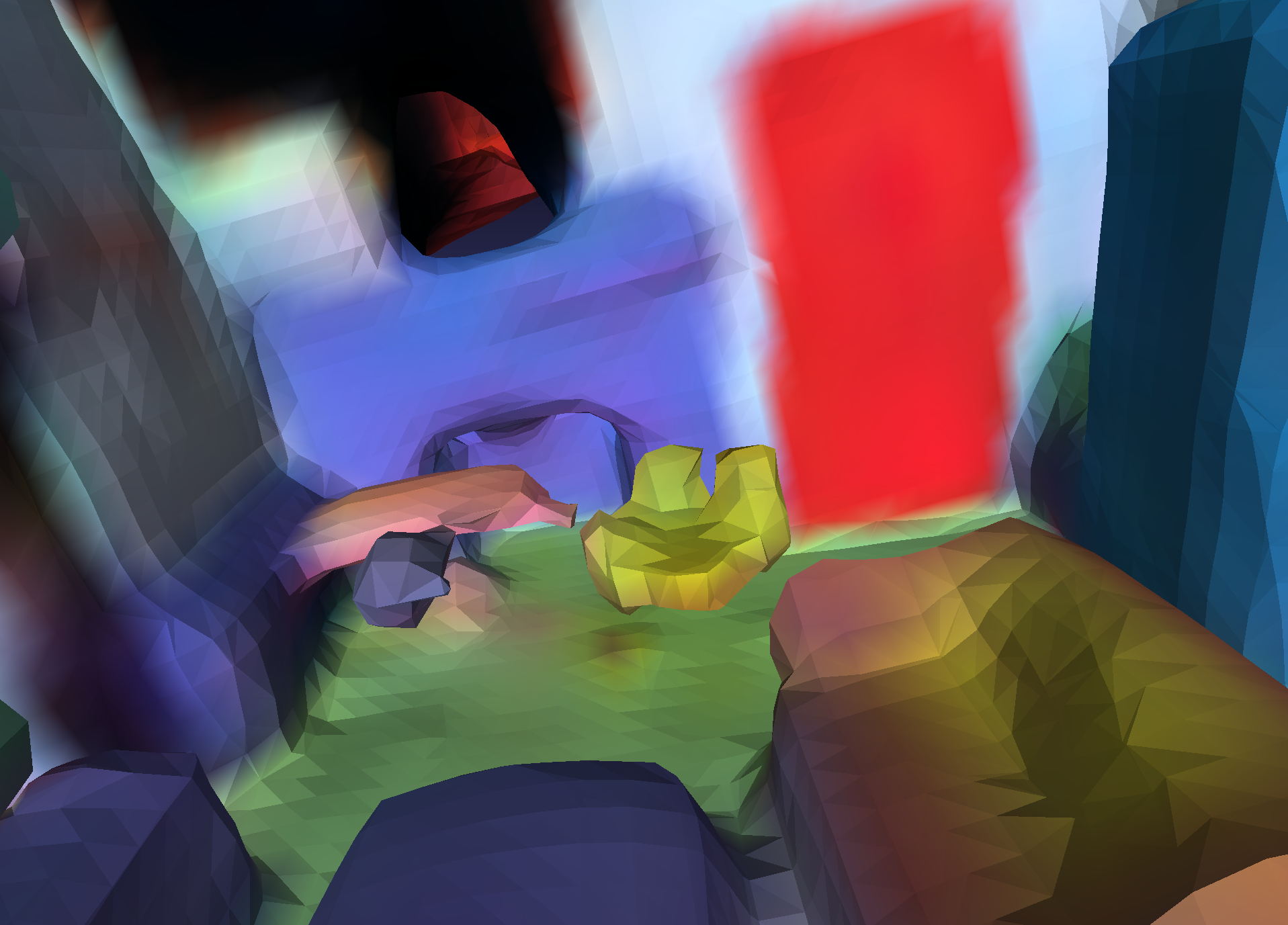}
            \caption{\scriptsize Low-Grained Rendering}
        \end{subfigure}
        \hfill % 使两个子图之间有一定间隔
        \begin{subfigure}[t]{0.48\textwidth}
            \centering
            \includegraphics[width=\textwidth]{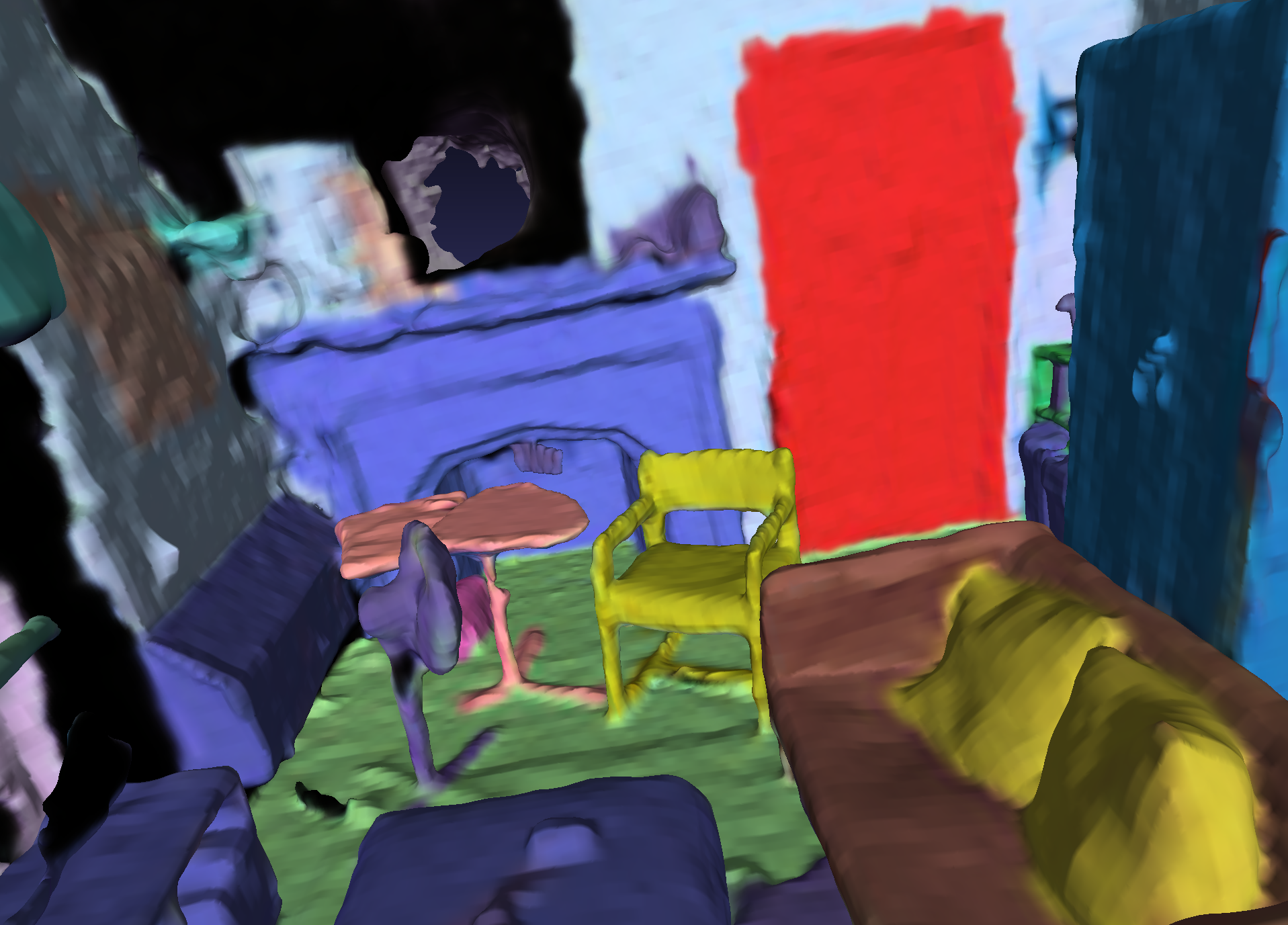}
            \caption{\scriptsize Fine-Grained Rendering}
        \end{subfigure}
    \end{minipage}
    \caption{Visualization of Low-Grained and Fine-Grained Rendering in PFPSMS.}
    \label{fig:rendering}
\end{figure}
\textbf{Pseudo-color Images:} Due to the lack of direct semantic information in the dataset, we employed the advanced DFormer \cite{Wang2023Dformer} to process the RGB images. During training, DFormer \cite{Wang2023Dformer} assigns a semantic label to each pixel. Then, using the label mapping mechanism from the NYU40 Dataset \cite{Silberman2012}, each semantic category is mapped to a specific color, generating pseudo-color images. Although these pseudo-color images represent semantic information through colors, they effectively capture the semantic context of the scene, providing a crucial semantic foundation for subsequent model training.
To improve the temporal efficiency of semantic modeling in hybrid scene representation, we employ the Progressive Semantic Modeling Strategy (PFPSMS). 

\textbf{Coarse-Grained Rendering:} In the initial phase, coarse-grained feature planes are used to perform rendering for half of the total iterations. During this stage, the voxel dimensions under coarse precision are set to 10 times larger than those used in the fine-precision phase for the latter half of the iterations. In this process, the ray weight $\omega_{\text{coarse}}(k)$ is calculated based on the volume density $\rho_k$ of the sampled points, using the NEUS method for coarse-grained rendering:
\begin{equation}
    \omega_{\text{coarse}}(k) = \exp\left(-\sum_{j=1}^{k-1} \sigma_j \Delta z_j\right) \cdot \left(1 - \exp(-\sigma_k \Delta z_k)\right)
\end{equation}
\begin{equation}
    \sigma_j = \varphi(sdf_j \cdot \text{inv}_s)
\end{equation}
where $\omega_{\text{coarse}}(k)$ represents the ray weight of the $k$-th voxel during the coarse-grained rendering phase, $\Delta z_k$ and $\Delta z_j$ are the distances between adjacent voxels under coarse precision, $\text{inv}_s$ is the scaling factor, $\varphi$ is the Sigmoid activation function, and $sdf_j$ is the SDF value of the $j$-th voxel.

Through computation, coarse-grained rendering rapidly captures the overall structure of the scene. Additionally, because the
number of voxels is relatively small, it significantly accelerates the rendering speed in the initial stages. This helps avoid unnecessary consumption of computational resources on fine details early on, while laying a foundation for more refined rendering in subsequent stages.

\textbf{Fine-Grained Rendering:} After completing half of the coarse-grained rendering iterations, the model switches to fine-grained rendering mode. In this phase, the voxel dimensions are restored to their standard size, as in traditional rendering. The number of voxels in fine-grained rendering is typically larger than in coarse-grained rendering, enabling the model to capture finer scene details and transformations. During fine-precision rendering, ray weights are computed based on the weights from the coarse-grained phase. The formula for calculating the ray weight $\omega_{\text{fine}}(k)$ in the fine-grained phase is as follows:
\begin{equation}
    \omega_{\text{fine}}(k) = \beta \cdot \omega_{\text{coarse}}(k) \cdot \frac{e^{-\sigma_k \Delta z_k}}{1 - \exp(-\sigma_{k+1} \Delta z_{k+1})}
\end{equation}
where $\beta$ is a pre-defined scaling factor. The symbol $\sigma_k$ represents the volume density of the sampled point $x_k$ corresponding to the $k$-th voxel. $\Delta z_k$ is the distance between two adjacent sampled points in the fine-precision rendering phase.

In the initial stage, the model quickly constructs a rough framework of the scene through coarse-grained rendering, accurately capturing the overall layout. Subsequently, fine-grained rendering focuses on the intricate details of the complex interactions between objects and the environment. This progressive strategy not only saves rendering time but also effectively avoids local optima, ensuring a balance between rendering efficiency and detail precision, thereby meeting the high-quality reconstruction requirements of complex indoor scenes.

\subsection{Optimization: Loss Function}

In the PreSem-Surf framework, we design a comprehensive loss function to jointly optimize rendering quality, geometric representation, and semantic information. We sample \( M \) pixels from the input images to define the overall loss function as follows:
\begin{equation}
    \mathcal{L} = \lambda_{\text{SG}} \mathcal{L}_{\text{SG}} + \lambda_{\text{sem}} \mathcal{L}_{\text{sem}}
\end{equation}
where \( \mathcal{L}_{\text{SG}} \) denotes the loss guided by the SG-MLP, and \( \mathcal{L}_{\text{sem}} \) is the semantic-modeling-guided loss.
\par
\textbf{SG-MLP Loss:}
\begin{equation}
	\begin{split}
		\mathcal{L}_{\text{SG}} = &\ \lambda_{\text{PR}} \mathcal{L}_{\text{PR}} + \lambda_{\text{rgb}} \mathcal{L}_{\text{rgb}} + \lambda_{d} \mathcal{L}_{d} \\
		&\ + \lambda_{\text{sdf}} \mathcal{L}_{\text{sdf}} + \lambda_{\text{eik}} \mathcal{L}_{\text{eik}} + \lambda_{\text{smooth}} \mathcal{L}_{\text{smooth}}
	\end{split}
\end{equation}
where $\lambda_{\text{PR}}, \lambda_{\text{rgb}}, \lambda_{d}, \lambda_{\text{sdf}}, \lambda_{\text{eik}}$ and $\lambda_{\text{smooth}}$ are the weights of different loss components.

\(\mathcal{L}_{\text{PR}}\) is the SDF loss guided by the PR MLP. Specifically, it is computed via uniform sampling in the truncated region and minimizes the difference between the predicted voxel distance and the ground-truth distance:
\begin{equation}
    \mathcal{L}_{\text{PR}} = \frac{1}{S_{\text{tr}}} \sum_{p \in S_{\text{tr}}} (D_p - \hat{D}_p)^2
\end{equation}
where \( S_{\text{tr}} \) is the set of sampled points in the truncated region, and \( D_p, \hat{D}_p \) represent the ground-truth and predicted distance values, respectively.

\(\mathcal{L}_{\text{rgb}}\) is sampled from all rays in space to measure the discrepancy between the rendered pixel color and the true pixel color:
\begin{equation}
	\mathcal{L}_{\text{rgb}} = \frac{1}{N_{\text{rgb}}} \sum_{m = 1}^{N_{\text{rgb}}} \ell_{\text{rgb}, m}
\end{equation}
where \(\mathcal{L}_{\text{rgb}, m}\) represents the RGB loss for the $m$-th sampled ray,and $N_{rgb}$ is the number of sampled rays.

\(\mathcal{L}_{d}\) is calculated based on rays with valid depth values to measure the difference between the rendered depth and the actual depth. Let $\mathcal{R}_{d}$ be the set of rays with valid depth values. For each ray $\mathcal{r} \in \mathcal{R}_{d}$, $l_{d}^{r}$ represents the difference between the rendered depth and the actual depth for ray $\mathcal{r}$:
\begin{equation}
    \mathcal{L}_{d} = \frac{1}{\lvert \mathcal{R}_{d} \rvert} \sum_{r \in \mathcal{R}_{d}} l_{d}^{r}
\end{equation}
\(\mathcal{L}_{\text{sdf}}\) and \(\mathcal{L}_{\text{fs}}\) are applied to disjoint sets of sample ray points. Specifically, \(l_{\text{sdf}}(x_{s})\) denotes the SDF loss for the sample point \(x_{s}\), while \(l_{\text{fs}}(x_{s})\) denotes the FS loss for that point:
\begin{equation}
    \mathcal{L}_{\text{sdf}} = \frac{1}{M} \sum_{m=1}^{M} \left( \frac{1}{\lvert S_{\text{tr}} \rvert} \sum_{s \in S_{\text{tr}}} l_{\text{sdf}}(x_{s}) \right)
\end{equation}
where $M$ represents the number of sampled rays with valid SDF values, and $S_{tr}$ is the set of sampled points in the truncated region.

\begin{equation}
    \mathcal{L}_{\text{fs}} = \frac{1}{M} \sum_{m=1}^{M} \left( \frac{1}{\lvert S_{\text{fs}} \rvert} \sum_{s \in S_{\text{fs}}} l_{\text{fs}}(x_{s}) \right)
\end{equation}
where $S_{fs}$ is the set of sampled rays with valid fs value.

\(\mathcal{L}_{\text{eik}}\) arises from the observation that points far from the reconstructed surface in space are often insufficiently constrained by the standard signed-distance function (SDF) loss. Therefore, we introduce \(\mathcal{L}_{\text{eik}}\) and apply it to \(S_{\text{fs}}\) to regularize those points so that they retain a valid signed-distance function (SDF) constraint. \(l_{\text{eik}}\bigl(x_{s}\bigr)\) denotes the individual Eikonal loss computed at the sample point:
\begin{equation}
	\mathcal{L}_{\text{eik}} = \frac{1}{M} \sum_{m = 1}^{M} \left( \frac{1}{\lvert S_{\text{fs}} \rvert} \sum_{s \in S_{\text{fs}}} l_{\text{eik}}\bigl(x_{s}\bigr) \right)
\end{equation}
\(\mathcal{L}_{\text{smooth}}\) ensures smoothness in the reconstructed result for points far from the surface. We introduce \(\mathcal{L}_{\text{smooth}}\) and apply this loss to near - surface points \(S_g\) randomly sampled over the entire voxel grid, enhancing the overall smoothness of the reconstruction. \(l_{\text{smooth}}(x_{s})\) represents the individual smoothing loss computed at the sample point:
\begin{equation}
	\mathcal{L}_{\text{smooth}} = \frac{1}{\lvert S_g \rvert} \sum_{s \in S_g} l_{\text{smooth}}(x_{s})
\end{equation}

In the semantic modeling process, we convert semantic labels into pseudo-color images for training. Consequently, the semantic loss can be split into two parts: one for the color loss of the pseudo-color images \(\mathcal{L}'_{\text{rgb}}\) and one for the depth loss \(\mathcal{L}'_d\) that guides semantic modeling:
\begin{equation}
	\mathcal{L}_{\text{sem}} = \lambda'_{\text{rgb}} \cdot \mathcal{L}'_{\text{rgb}} + \lambda'_d \cdot \mathcal{L}'_d
\end{equation}

\section{Results}
\subsection{Experimental Setup}

\textbf{Datasets.} We conducted a quantitative evaluation of our method on 7 scenes from the Synthetic Dataset \cite{Azinovic2022}. To simulate the effects of real depth sensors, we added noise and artifacts to the rendered depth images.

\textbf{Evaluation Metrics.} This paper employs C-L1, NC, F-score, IoU, Acc, and Comp as the six metrics to comprehensively evaluate the reconstruction performance of various methods.

\textbf{Baseline.} Our method takes GO-Surf \cite{Wang2022go} as the baseline and is compared with the latest benchmark in the field and other state-of-the-art 3D reconstruction models, Neural\_RGBD \cite{Azinovic2022} and Co-slam \cite{Wang2023}.

\textbf{Experimental Setup.} All methods in this paper were tested on a desktop with an AMD Ryzen 9 5900X CPU with a base clock frequency of 3.7 GHz and an NVIDIA 4090 GPU.

The voxel sizes for fine sampling were set to \([0.03m, 0.06m, 0.24m, 0.96m]\), while for coarse sampling, the sizes were increased by a factor of 10. Optimization was performed in PyTorch using the Adam optimizer, with the learning rate set to 0.001 for each of the components: \(\text{NeRF}\), \(\text{SG-MLP}\), and \(\text{Segment-odel}\). The loss function weights were chosen as \(\lambda_{\text{model}} = 5\), \(\lambda_{\text{SG}} = 4\), and \(\lambda_{\text{sem}} = 1\).

\begin{figure*}[!t]
    \centering
    \scriptsize  % 可选: 调整整体字体，以节省空间
    % 第一行
    \begin{subfigure}[t]{0.22\textwidth}
        \centering
        \includegraphics[width=\textwidth]{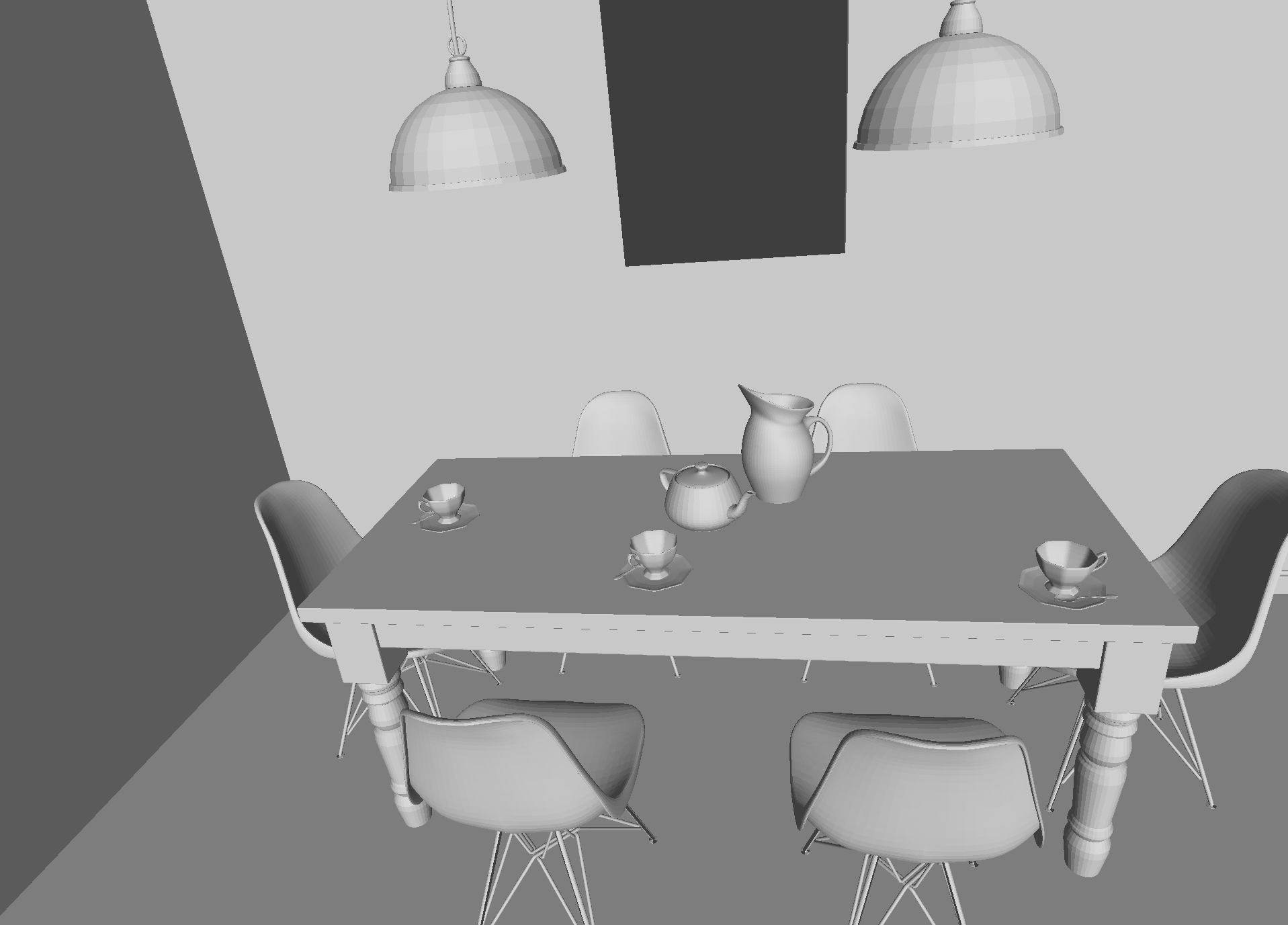}
        % 如果不需要单独标题，则去掉 \caption*
        % \caption*{\scriptsize Ground Truth}
    \end{subfigure}
    \begin{subfigure}[t]{0.22\textwidth}
        \centering
        \includegraphics[width=\textwidth]{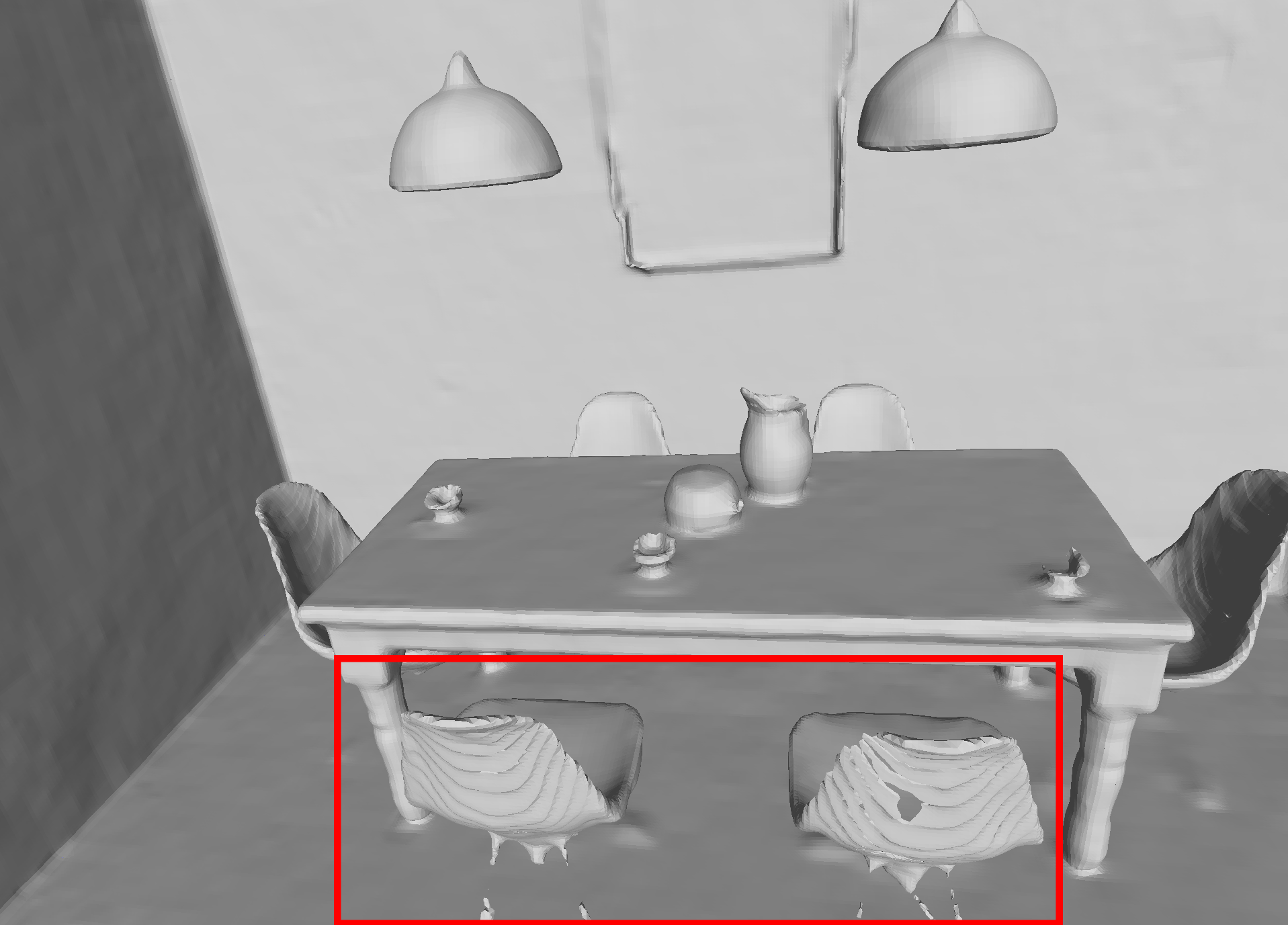}
        % \caption*{\scriptsize GO-Surf}
    \end{subfigure}
    \begin{subfigure}[t]{0.22\textwidth}
        \centering
        \includegraphics[width=\textwidth]{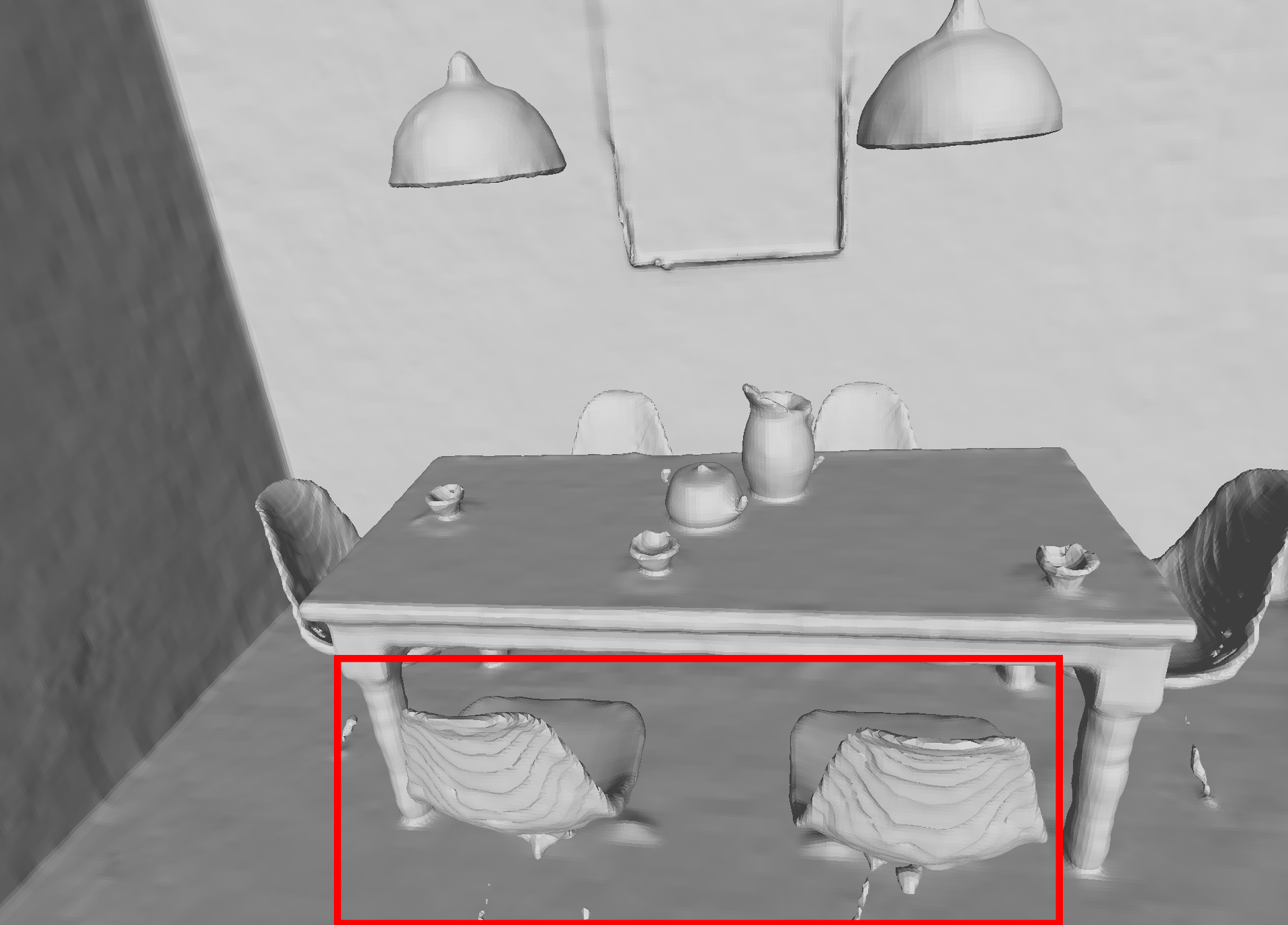}
        % \caption*{\scriptsize PreSem-Surf}
    \end{subfigure}
    \begin{subfigure}[t]{0.22\textwidth}
        \centering
        \includegraphics[width=\textwidth]{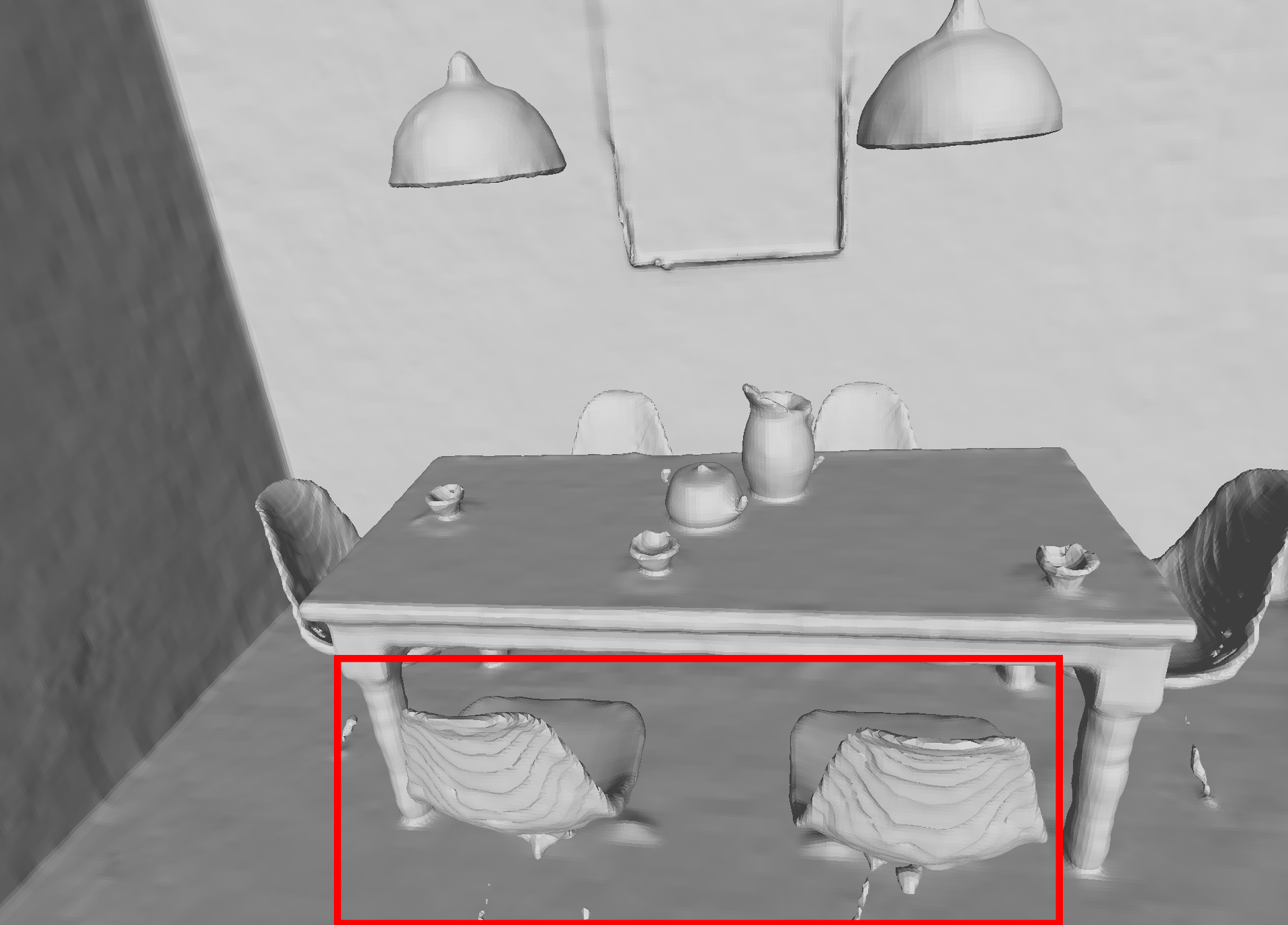}
        % \caption*{\scriptsize Neural-RGBD}
    \end{subfigure}
    % 换行
    \par\vspace{0.5em}
    % 第二行
    \begin{subfigure}[t]{0.22\textwidth}
        \centering
        \includegraphics[width=\textwidth]{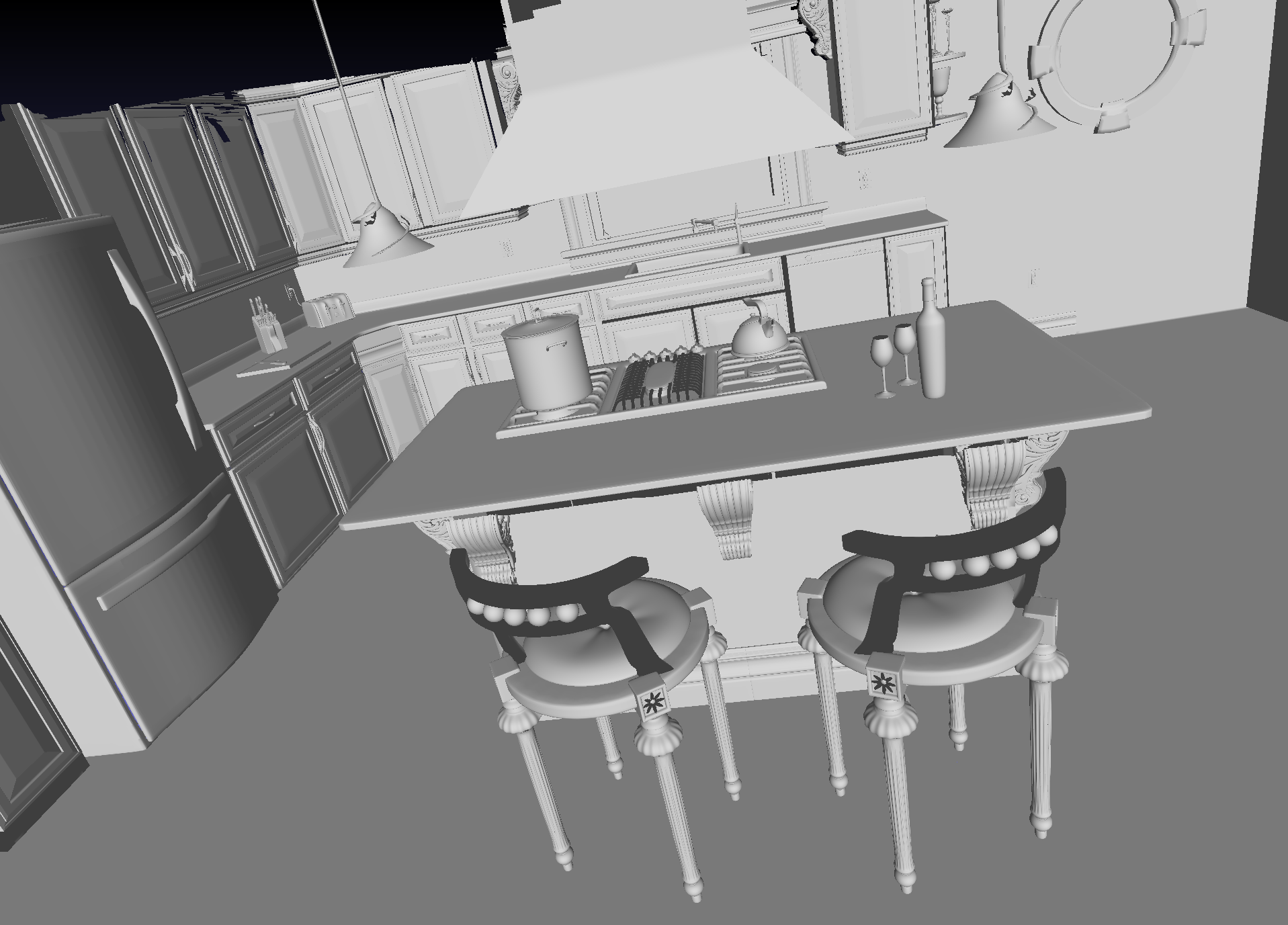}
    \end{subfigure}
    \begin{subfigure}[t]{0.22\textwidth}
        \centering
        \includegraphics[width=\textwidth]{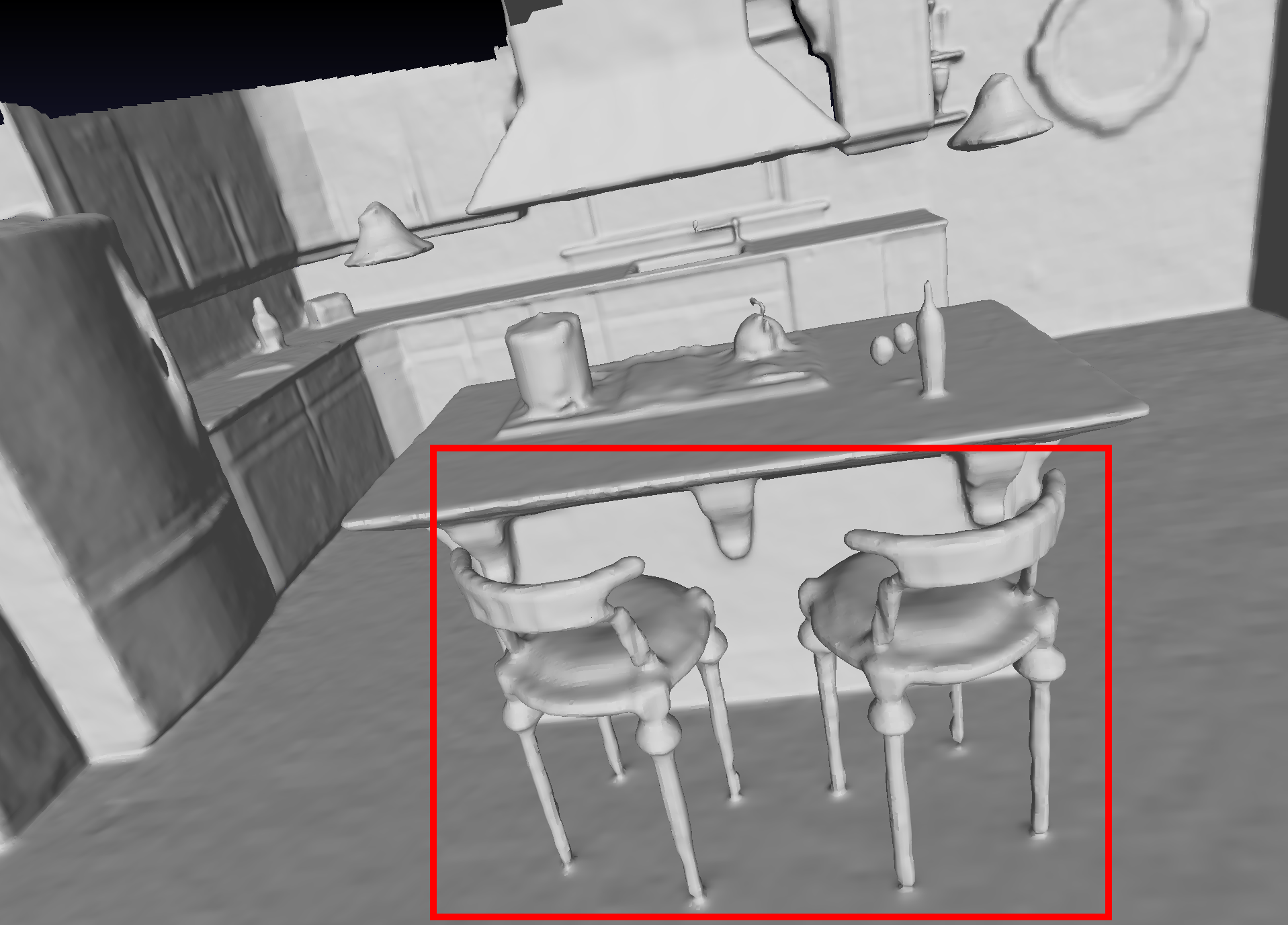}
    \end{subfigure}
    \begin{subfigure}[t]{0.22\textwidth}
        \centering
        \includegraphics[width=\textwidth]{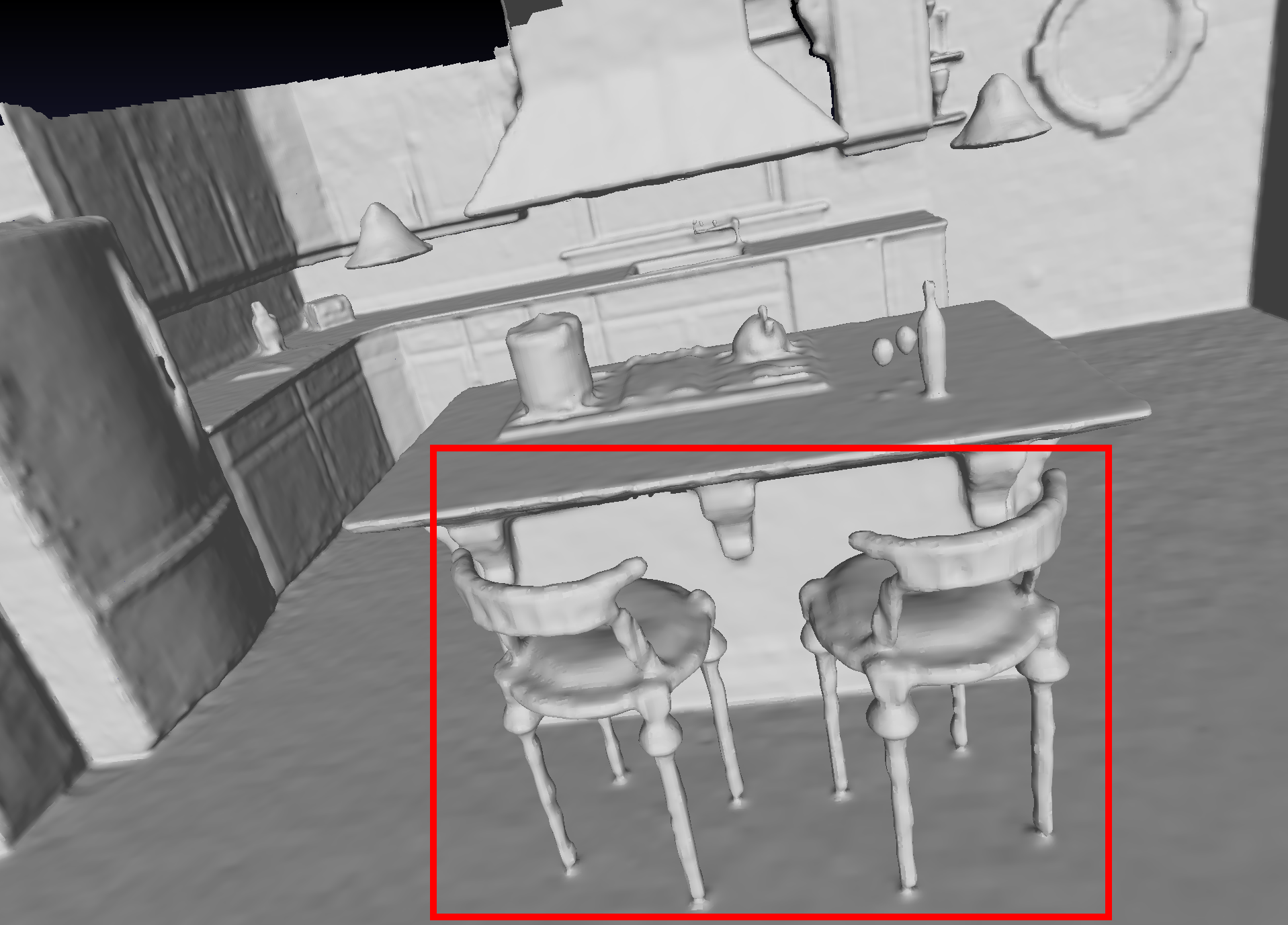}
    \end{subfigure}
    \begin{subfigure}[t]{0.22\textwidth}
        \centering
        \includegraphics[width=\textwidth]{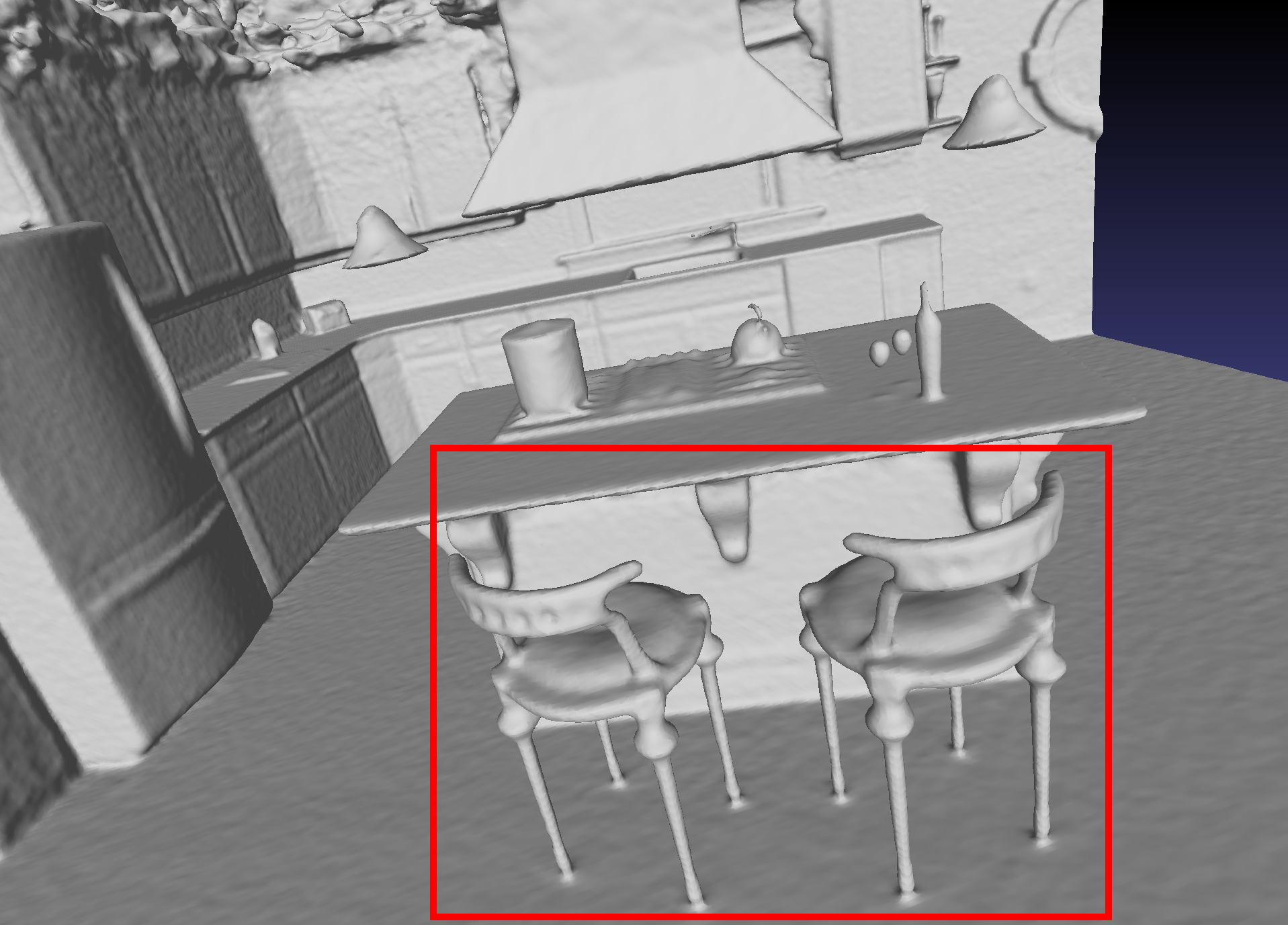}
    \end{subfigure}
    % 第三行
    \par\vspace{0.5em}
    \begin{subfigure}[t]{0.22\textwidth}
        \centering
        \includegraphics[width=\textwidth]{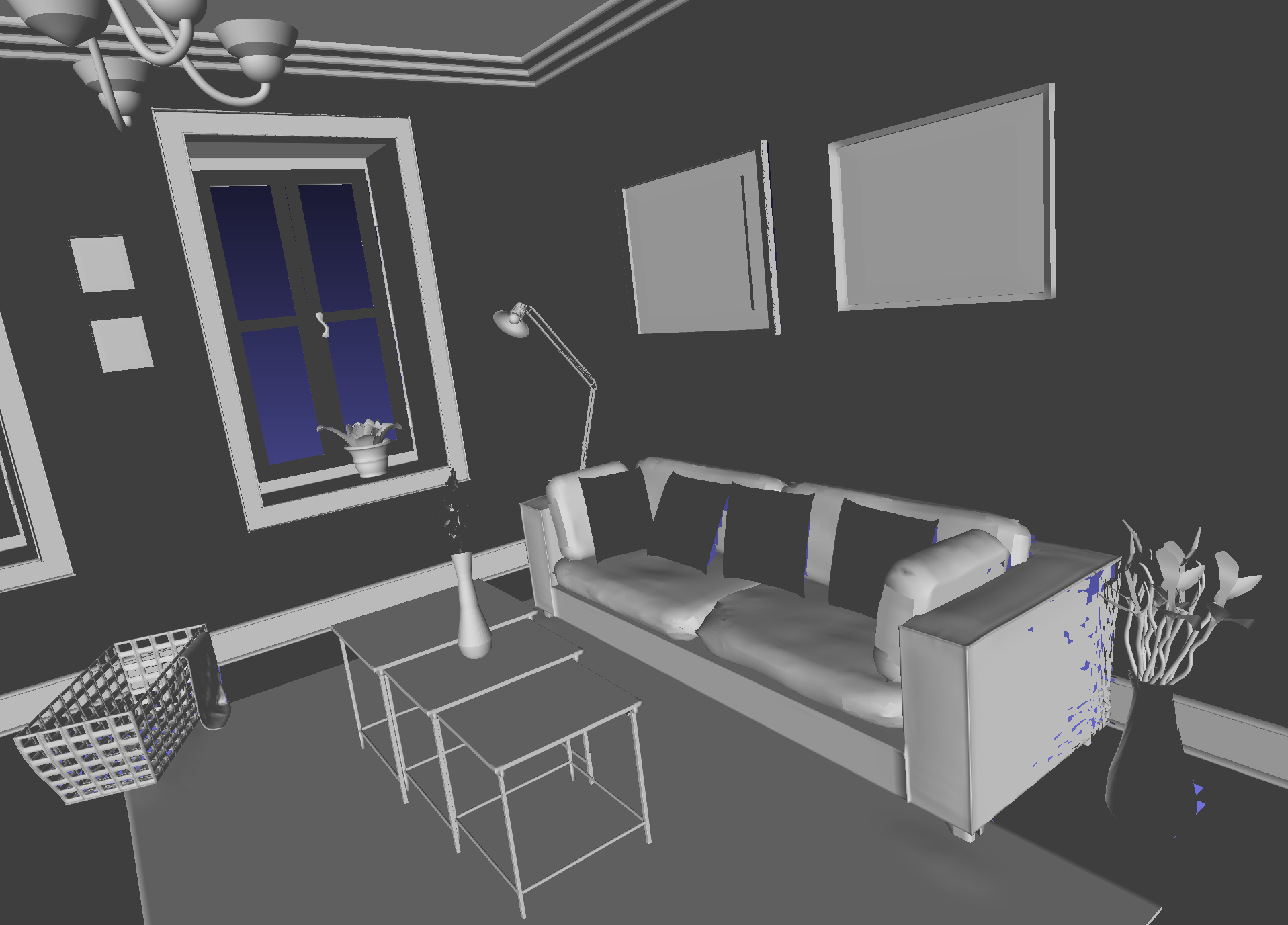}
    \end{subfigure}
    \begin{subfigure}[t]{0.22\textwidth}
        \centering
        \includegraphics[width=\textwidth]{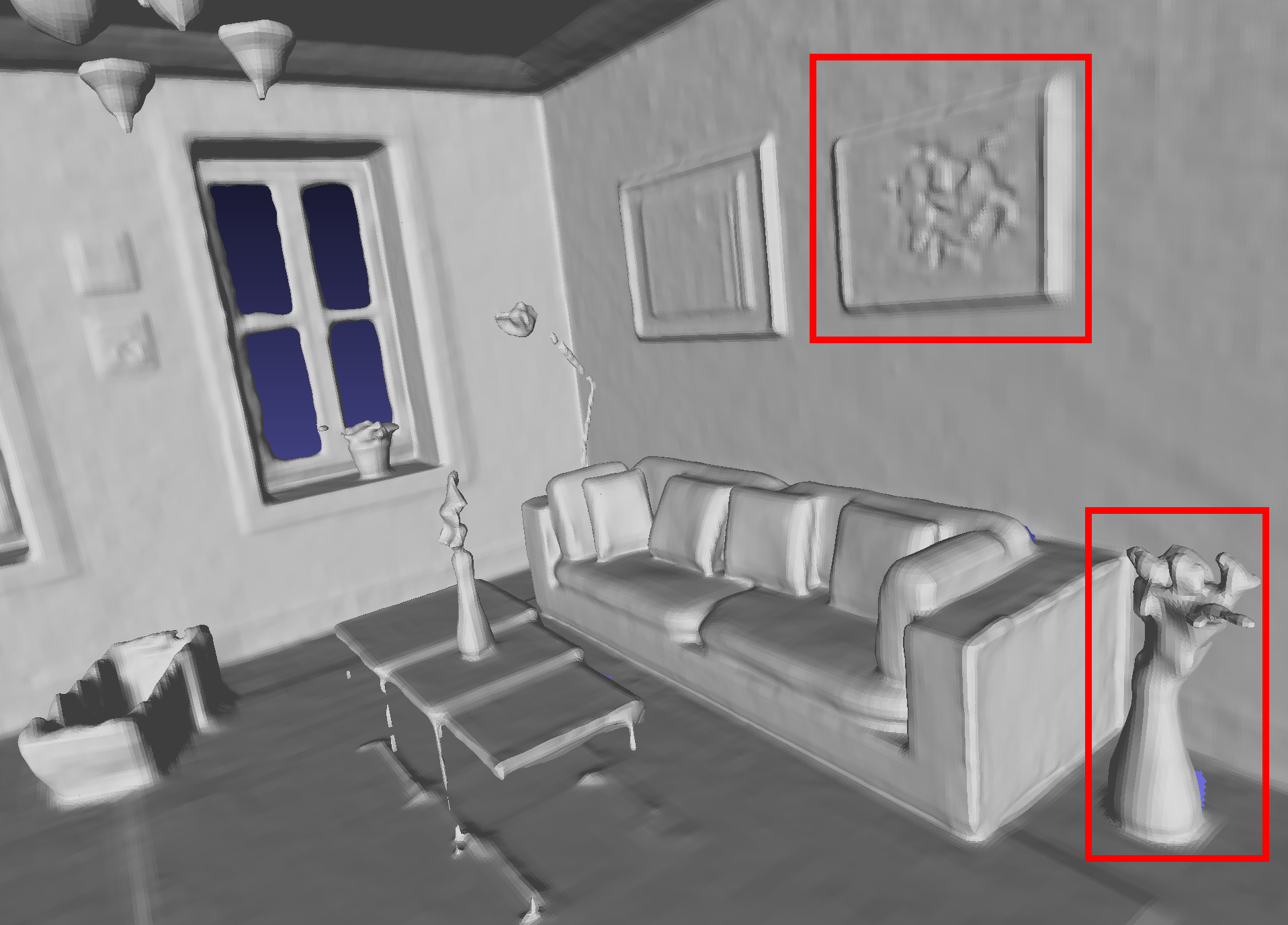}
    \end{subfigure}
    \begin{subfigure}[t]{0.22\textwidth}
        \centering
        \includegraphics[width=\textwidth]{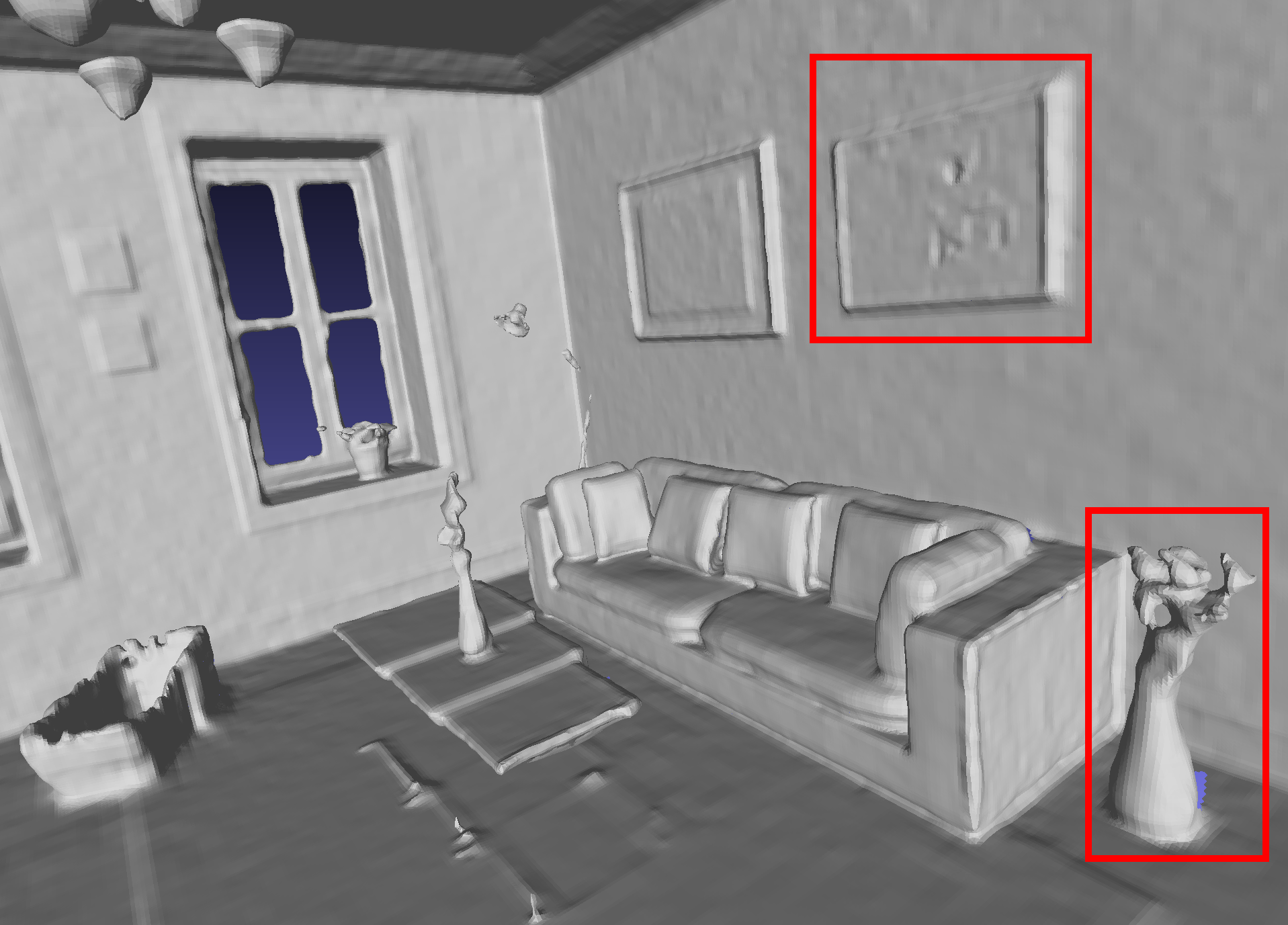}
    \end{subfigure}
    \begin{subfigure}[t]{0.22\textwidth}
        \centering
        \includegraphics[width=\textwidth]{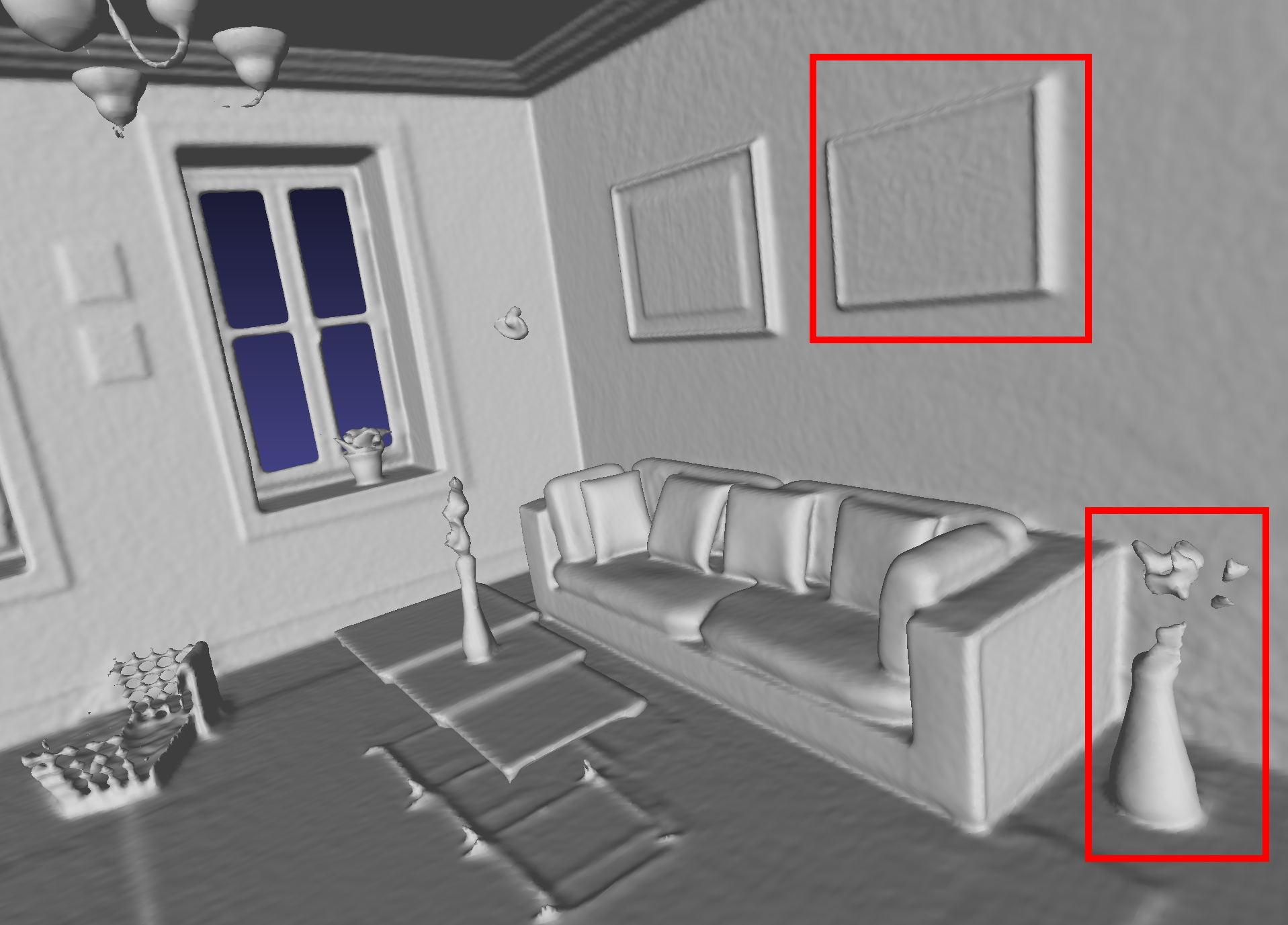}
    \end{subfigure}
    % grey_white_room
    \par\vspace{0.5em}
    \begin{subfigure}[t]{0.22\textwidth}
        \centering
        \includegraphics[width=\textwidth]{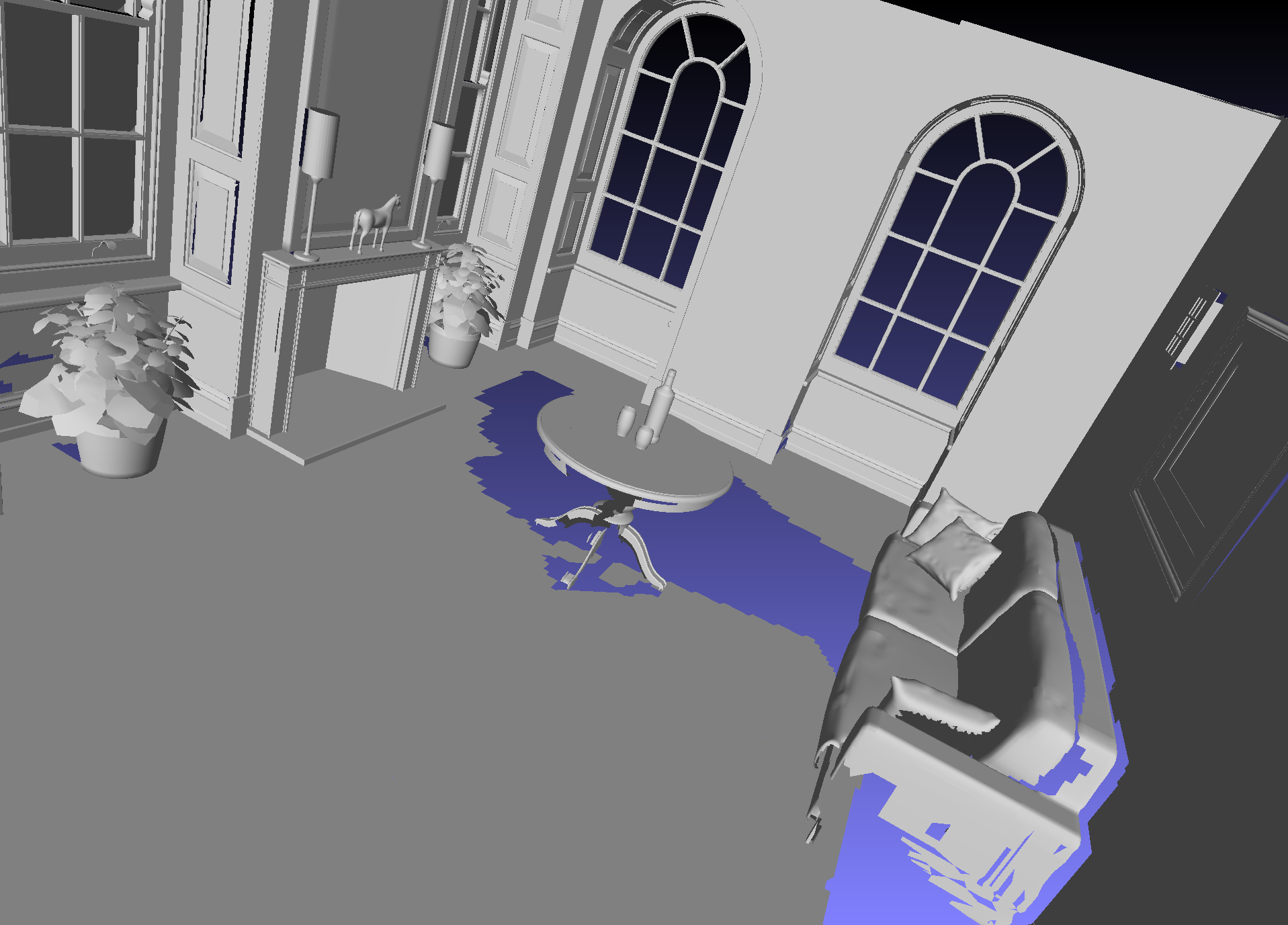}
    \end{subfigure}
    \begin{subfigure}[t]{0.22\textwidth}
        \centering
        \includegraphics[width=\textwidth]{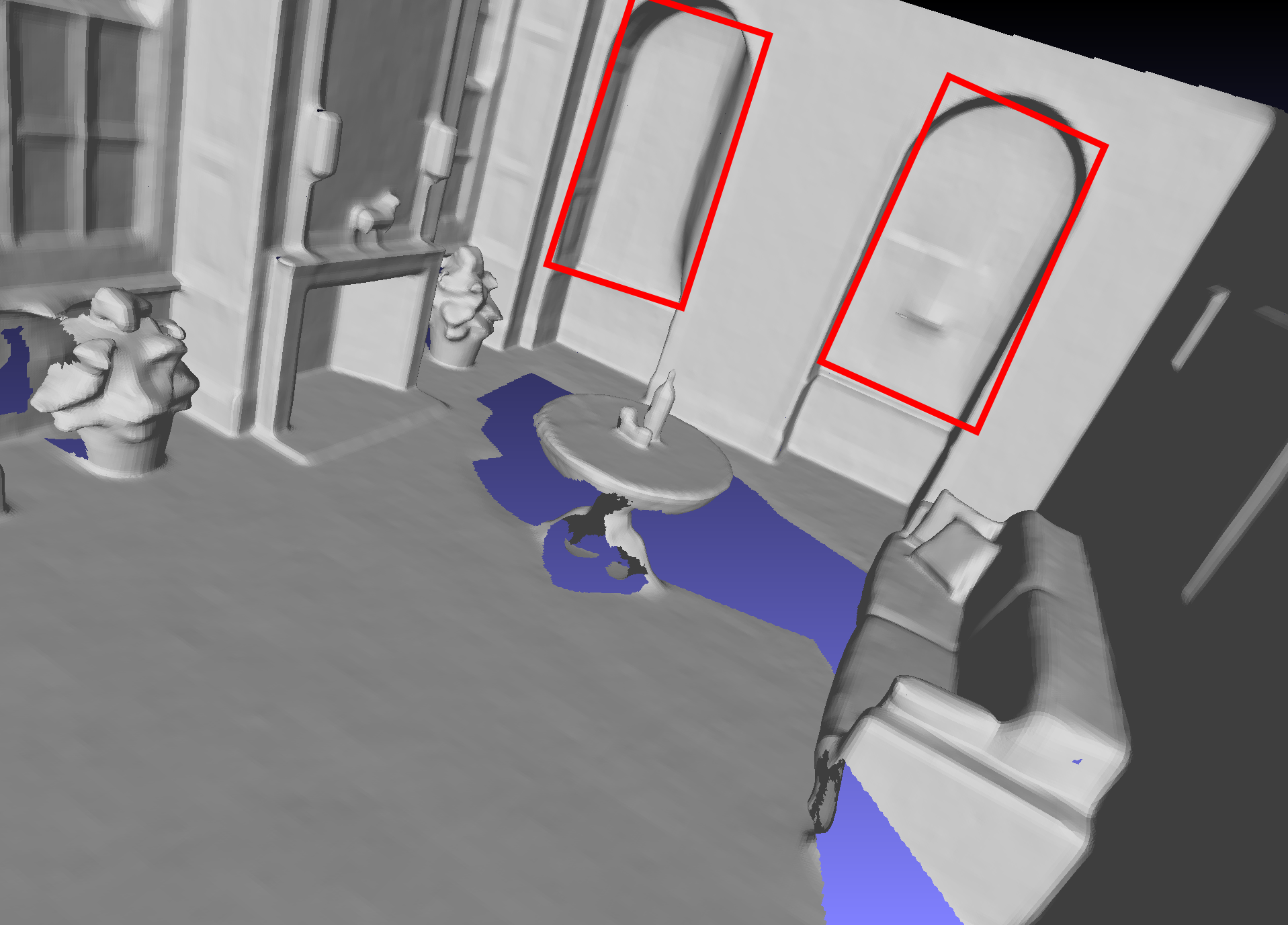}
    \end{subfigure}
    \begin{subfigure}[t]{0.22\textwidth}
        \centering
        \includegraphics[width=\textwidth]{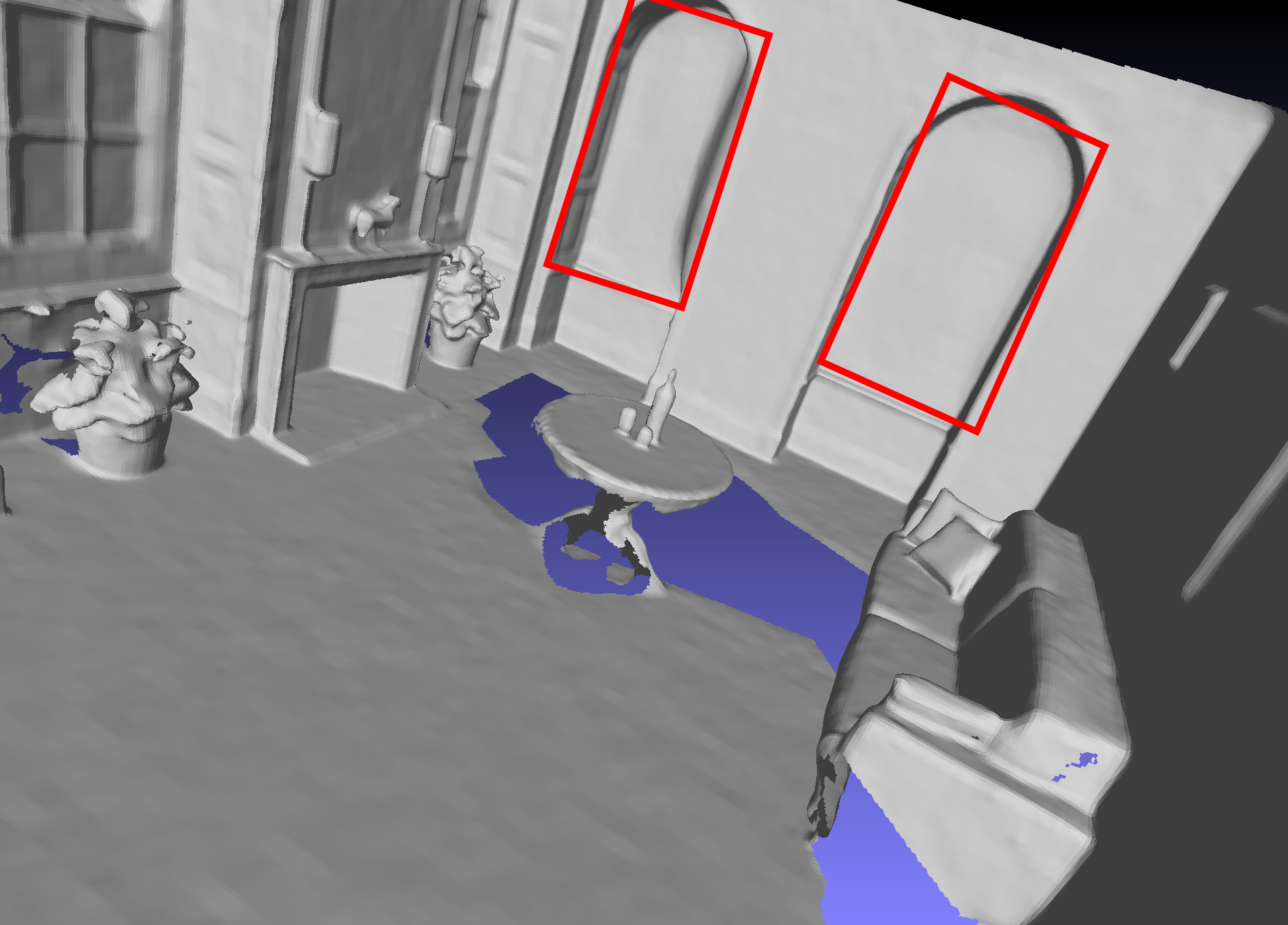}
    \end{subfigure}
    \begin{subfigure}[t]{0.22\textwidth}
        \centering
        \includegraphics[width=\textwidth]{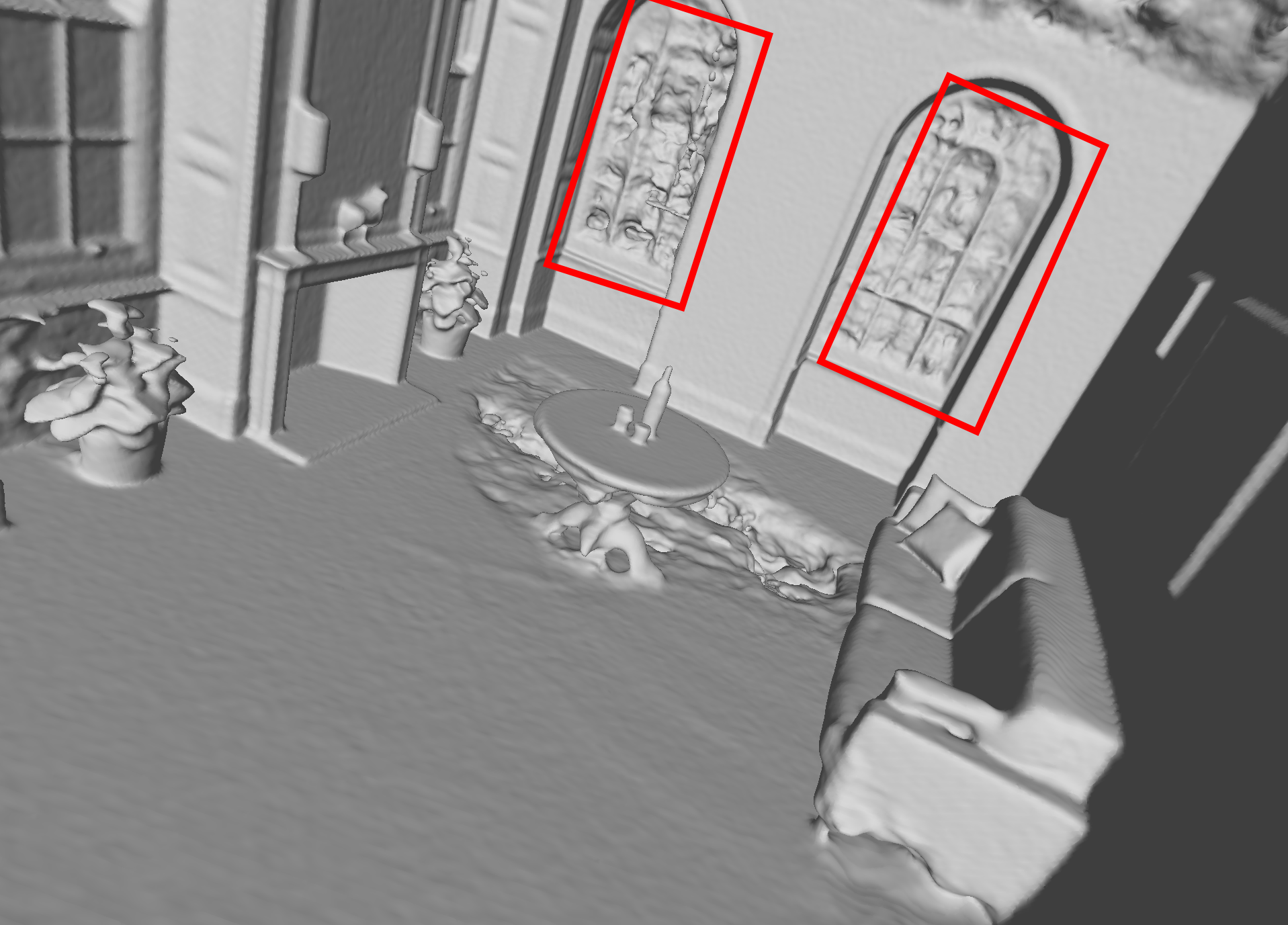}
    \end{subfigure}
    % thin_geometry
    \par\vspace{0.5em}
    \begin{subfigure}[t]{0.22\textwidth}
        \centering
        \includegraphics[width=\textwidth]{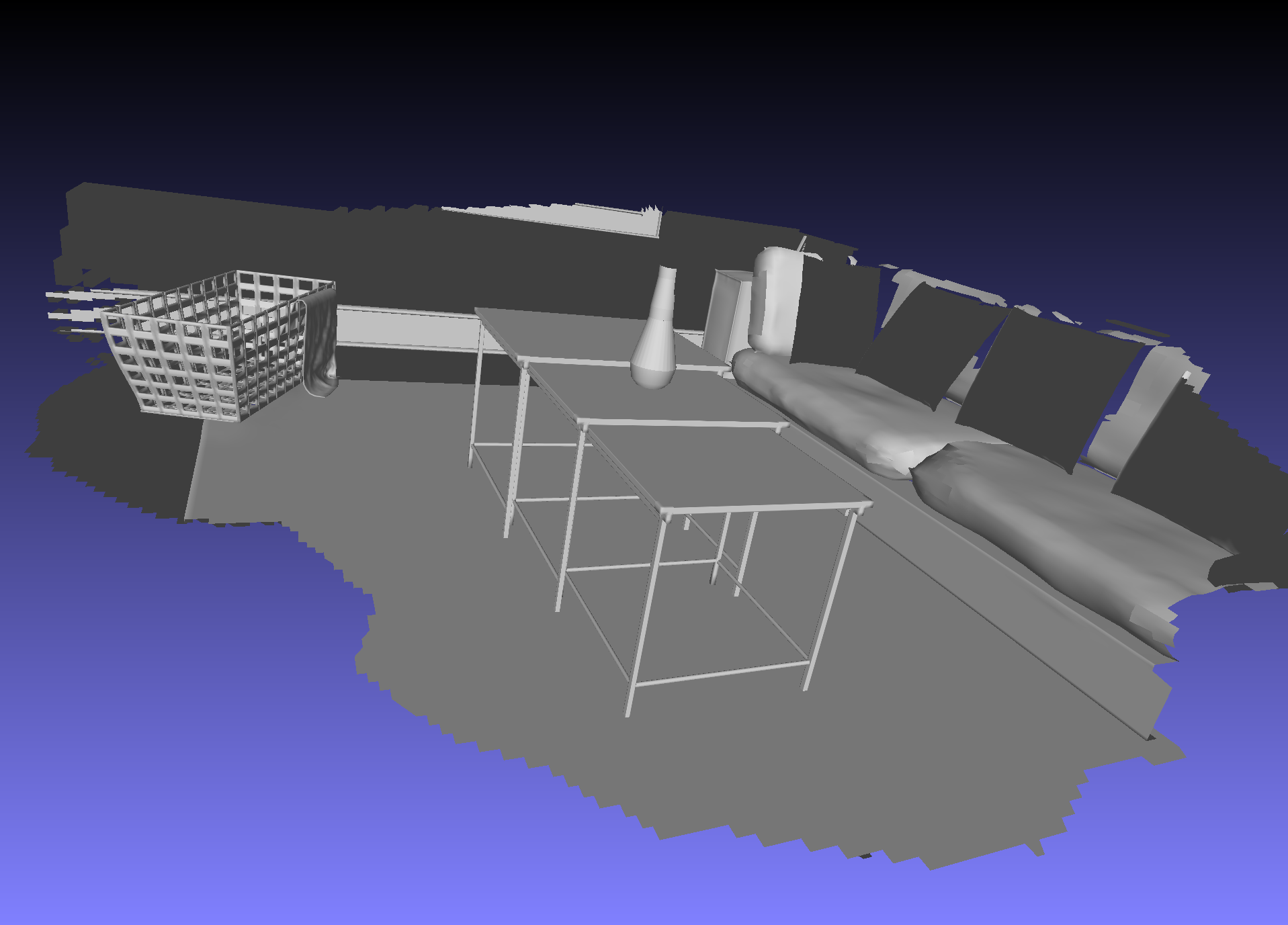}
    \end{subfigure}
    \begin{subfigure}[t]{0.22\textwidth}
        \centering
        \includegraphics[width=\textwidth]{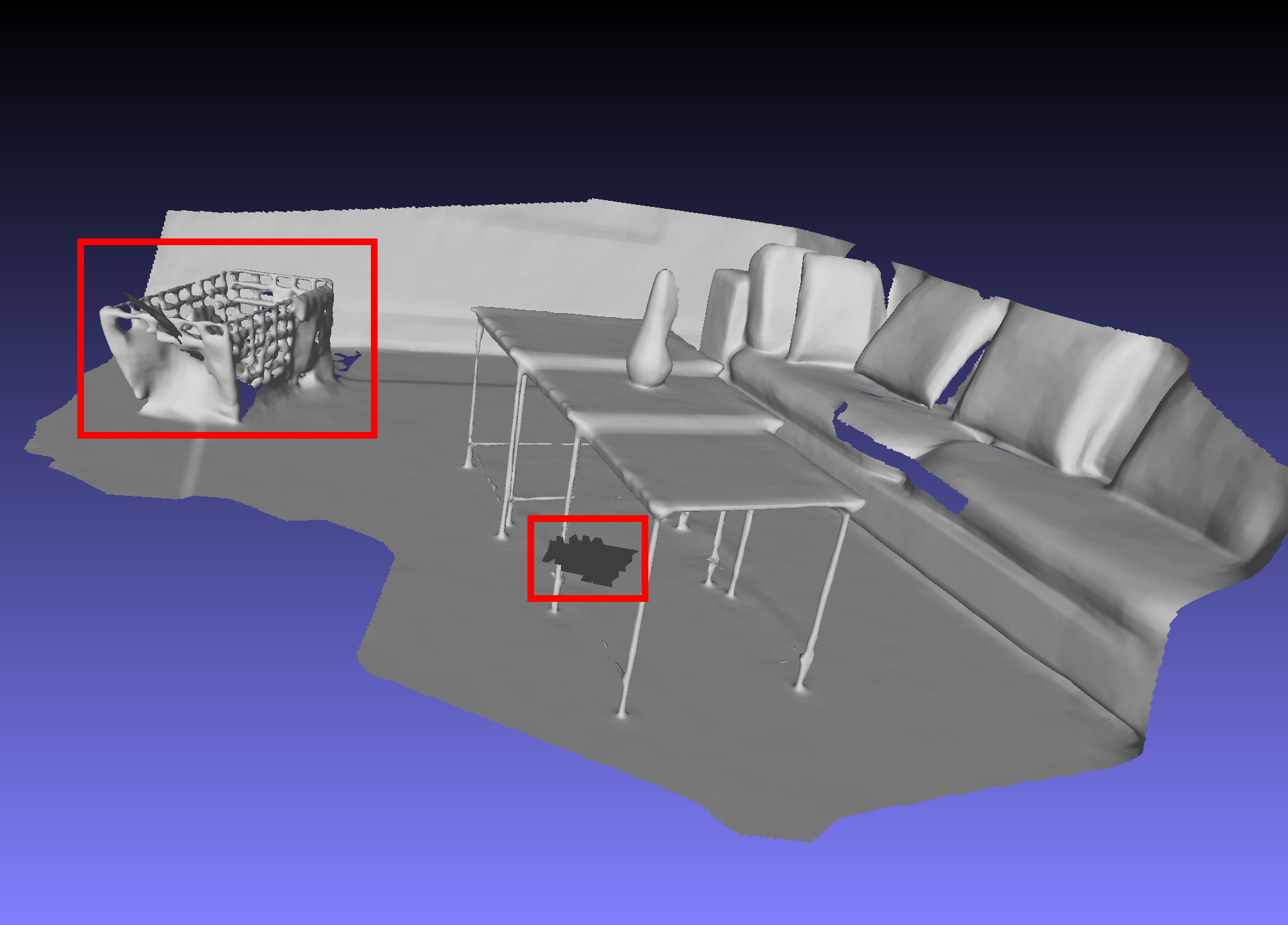}
    \end{subfigure}
    \begin{subfigure}[t]{0.22\textwidth}
        \centering
        \includegraphics[width=\textwidth]{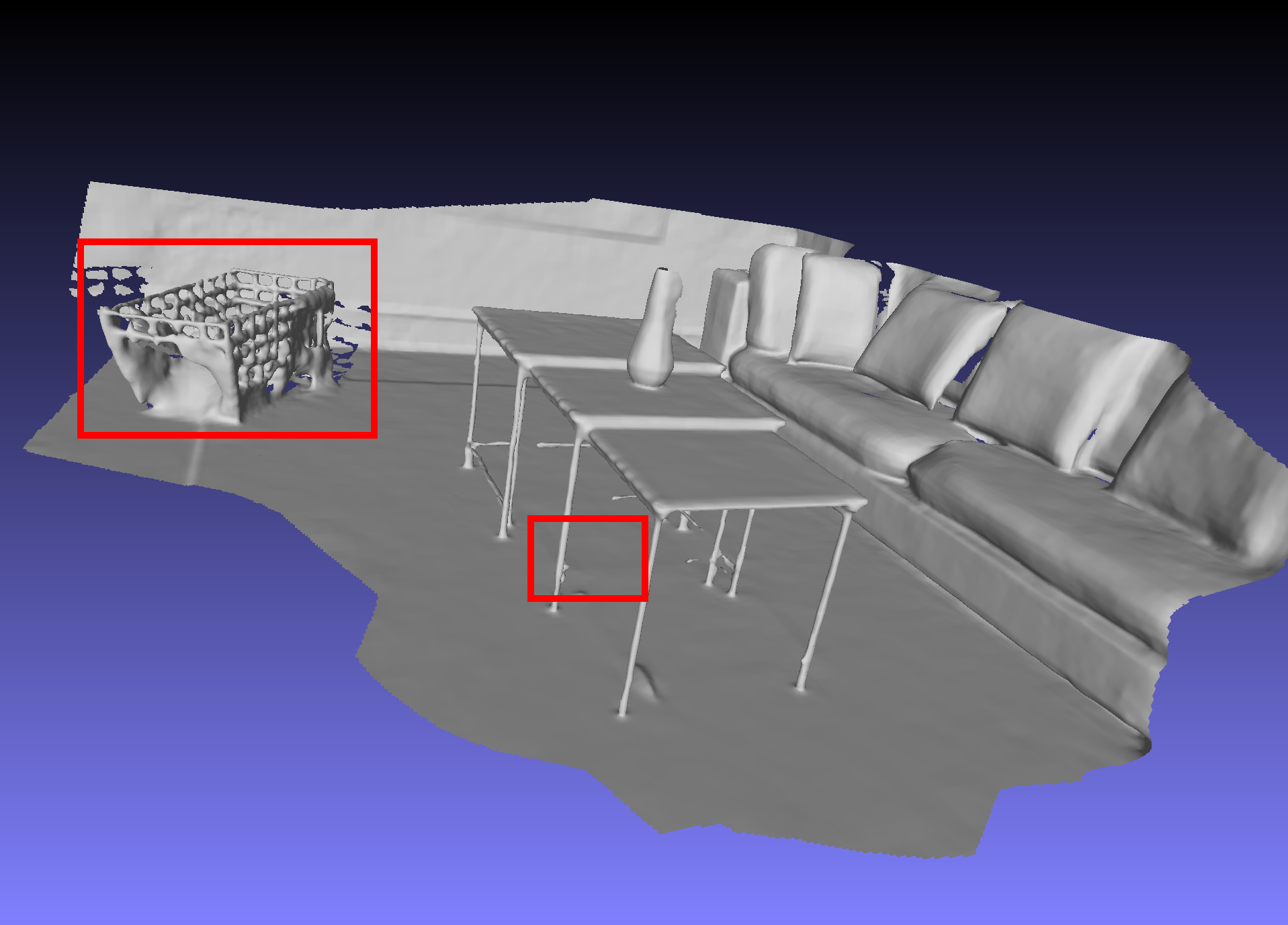}
    \end{subfigure}
    \begin{subfigure}[t]{0.22\textwidth}
        \centering
        \includegraphics[width=\textwidth]{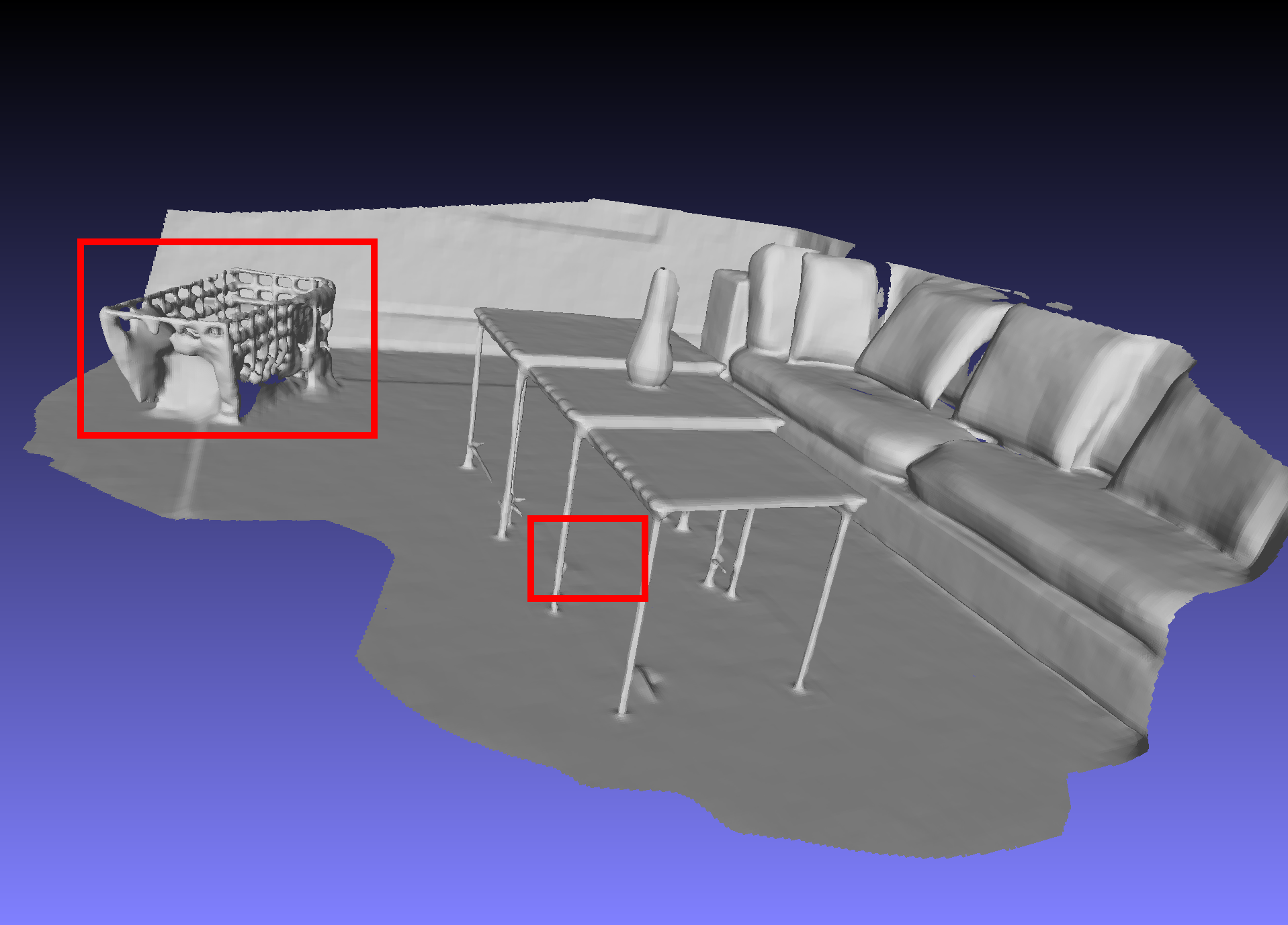}
    \end{subfigure}
    % whiteroom
    \par\vspace{0.5em}
    \begin{subfigure}[t]{0.22\textwidth}
        \centering
        \includegraphics[width=\textwidth]{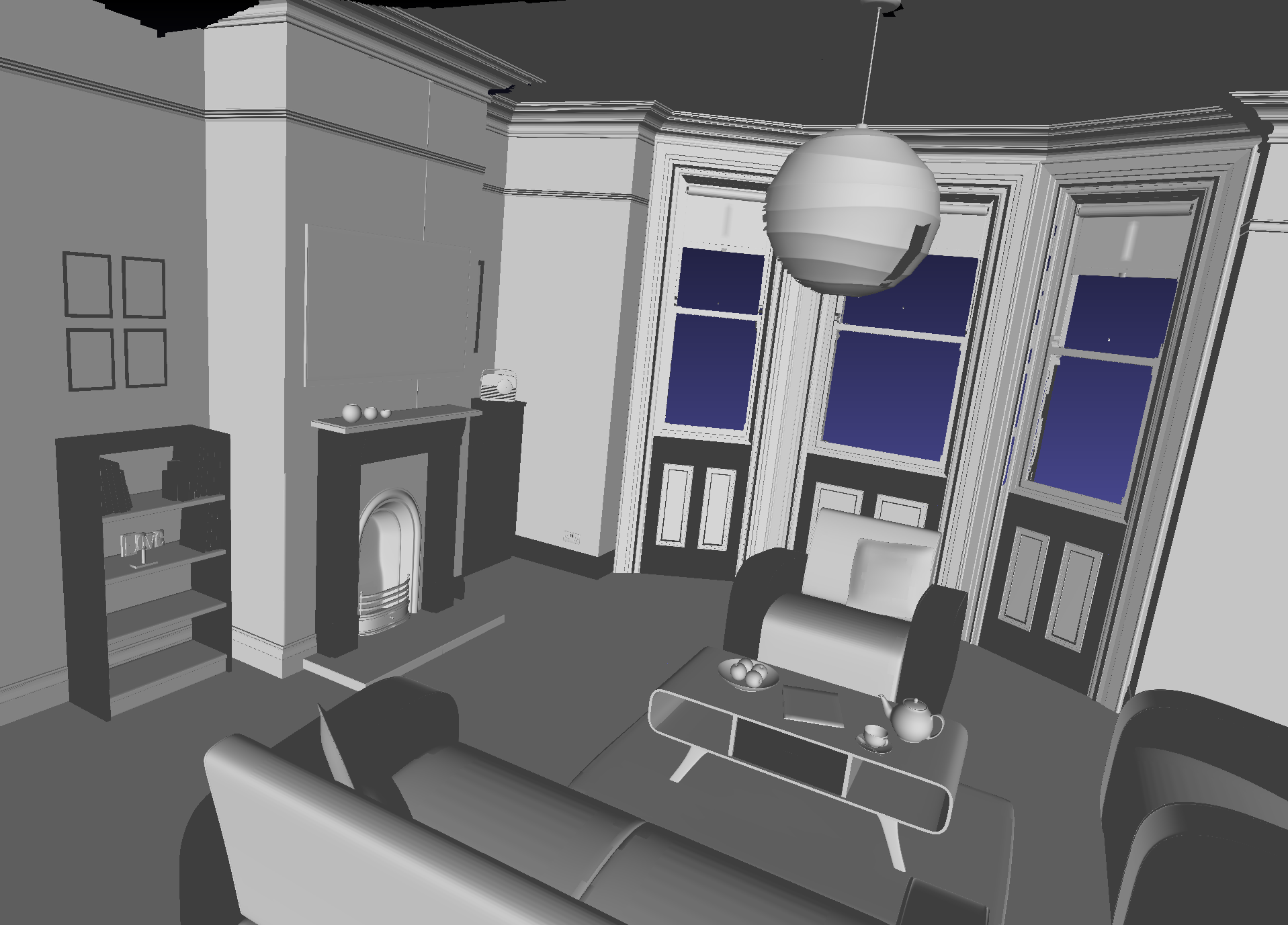}
        \caption*{\scriptsize Ground Truth}
    \end{subfigure}
    \begin{subfigure}[t]{0.22\textwidth}
        \centering
        \includegraphics[width=\textwidth]{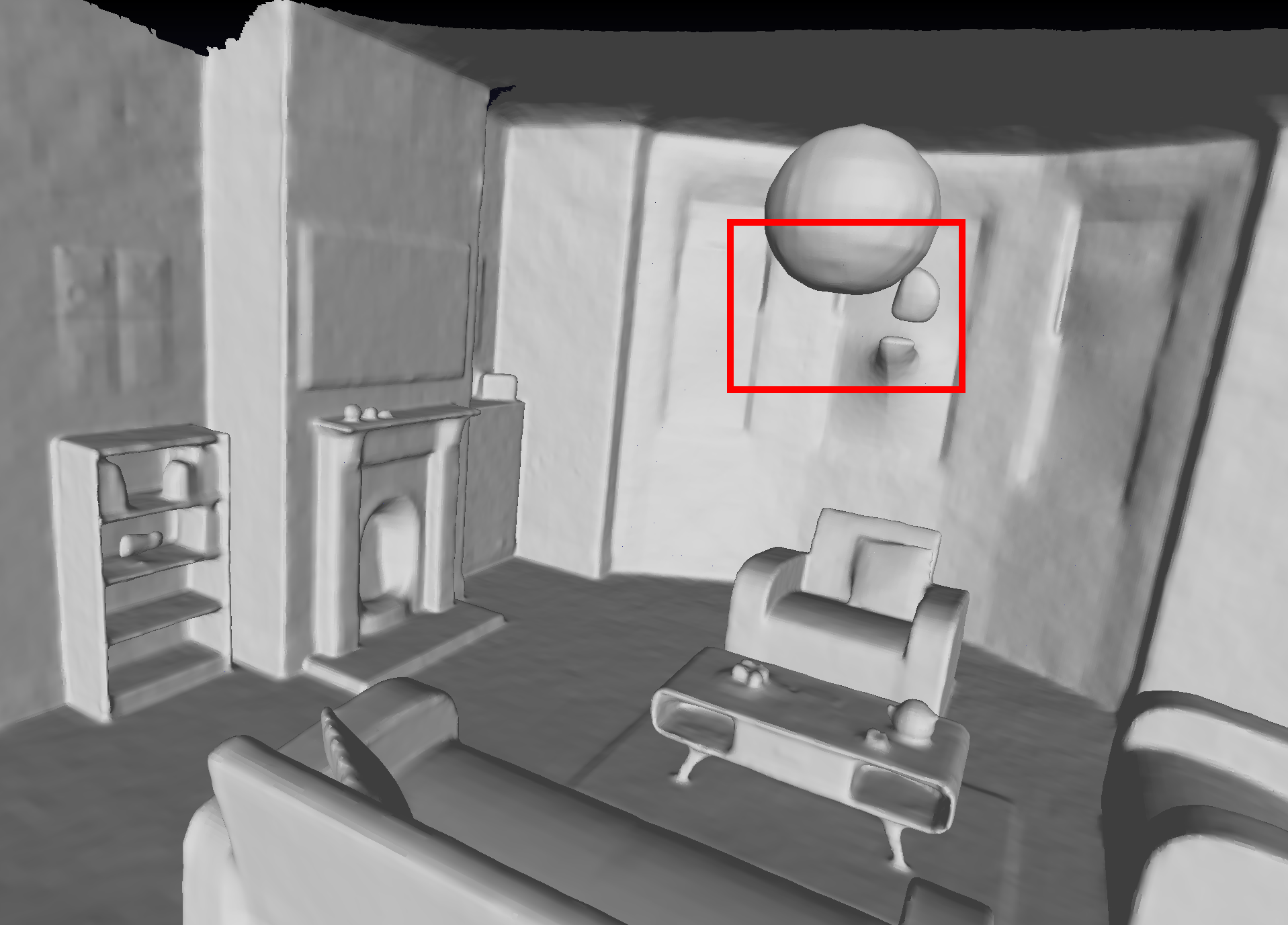}
        \caption*{\scriptsize GO-Surf}
    \end{subfigure}
    \begin{subfigure}[t]{0.22\textwidth}
        \centering
        \includegraphics[width=\textwidth]{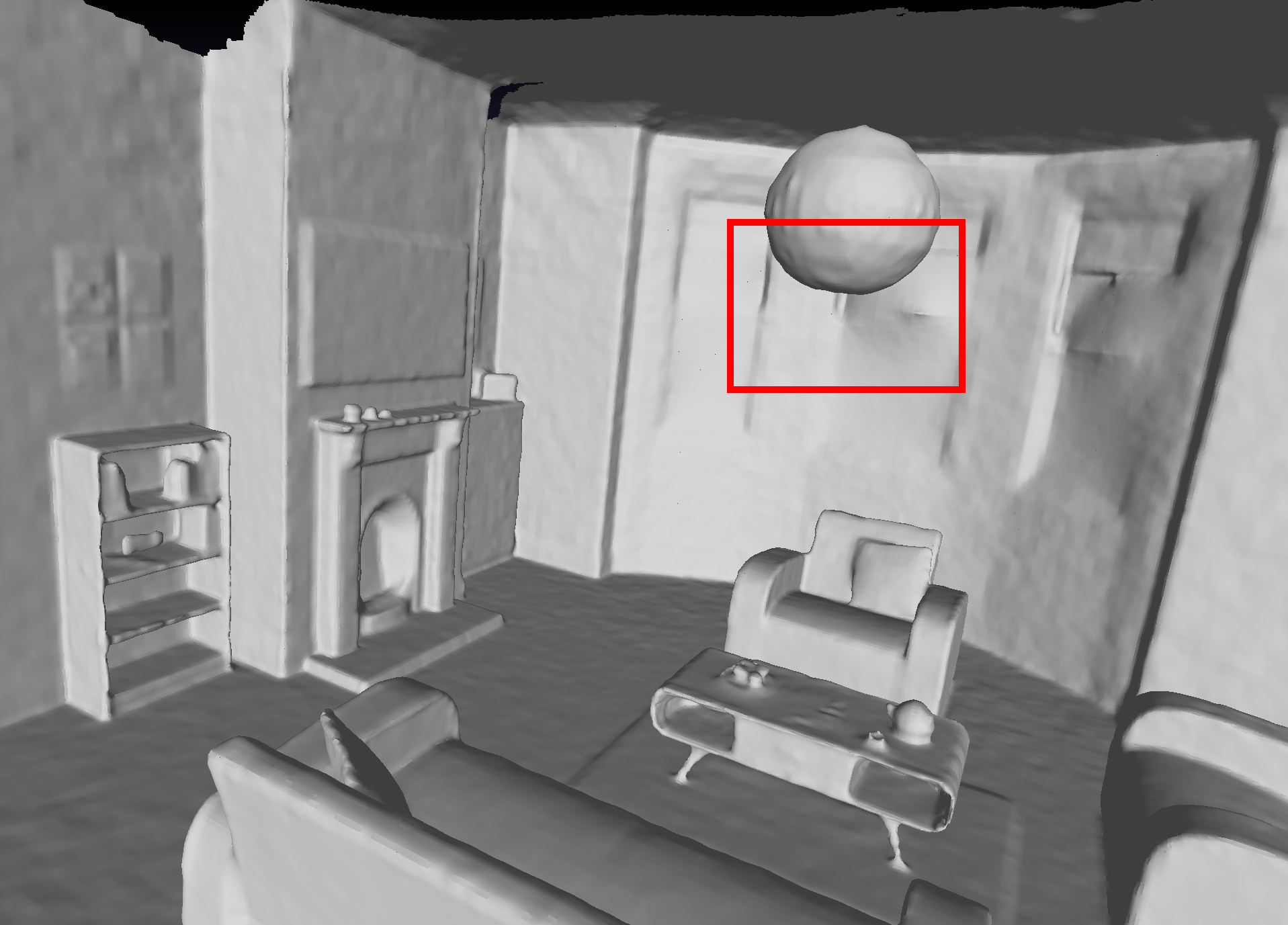}
        \caption*{\scriptsize PreSem-Surf}
    \end{subfigure}
    \begin{subfigure}[t]{0.22\textwidth}
        \centering
        \includegraphics[width=\textwidth]{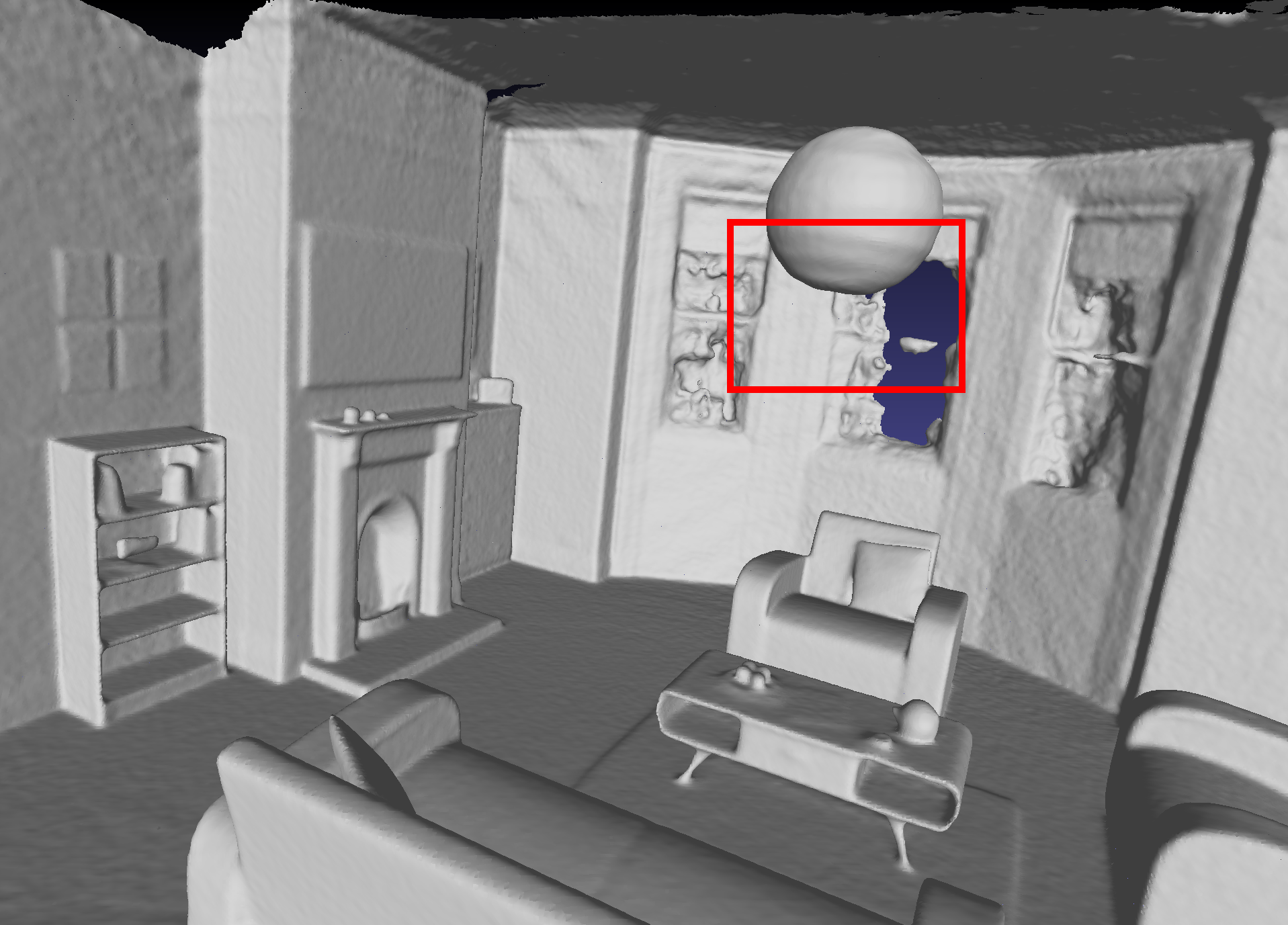}
        \caption*{\scriptsize Neural-RGBD}
    \end{subfigure}
    \caption{In the qualitative comparison of PreSem-Surf with baseline, we conducted a visual analysis on 6 scenes in
    the Synthetic Dataset and highlighted details with red squares. PreSem-Surf achieved better precision and smoothness
    in reconstruction.
    }
    \label{fig:qualitative-comparison}
\end{figure*}

\begin{table*}[!t]
    \centering
    \caption{Performance comparison across methods.}
    % 将整张表格缩放到 0.8倍 \textwidth 宽度
    \resizebox{0.8\textwidth}{!}{%
    \begin{tabular}{lcccccc}
        \hline
        \textbf{Model} & \textbf{C-L1 $\downarrow$} & \textbf{NC $\uparrow$} & \textbf{F-score $\uparrow$} & \textbf{IoU $\uparrow$} & \textbf{Acc $\downarrow$} & \textbf{Comp $\downarrow$} \\
        \hline
        Co-slam             & 0.0573 & 0.8904 & 0.8827 & 0.5459 & 0.0272 & \textbf{0.0246} \\
        NeRF-SLAM-Benchmark & 0.0836 & 0.8639 & 0.8375 & 0.4776 & 0.0327 & 0.0395 \\
        Neural RGBD         & \underline{0.0261} & 0.8993 & 0.9314 & \underline{0.5938} & \textbf{0.0191} & 0.0315 \\
        GO-Surf             & 0.0264 & \textbf{0.9138} & \underline{0.9329} & 0.5850 & 0.0210 & 0.0299 \\
        PreSem-Surf (Ours)  & \textbf{0.0236} & \underline{0.9132} & \textbf{0.9440} & \textbf{0.6389} & \underline{0.0204} & \underline{0.0250} \\
        \hline
    \end{tabular}%
    }
    \label{tab:performanceCompa}
\end{table*}
\begin{table}[t]
    \centering
    \caption{The performance metrics of PreSem-Surf in different scenarios.}
    \begin{tabular}{lcccccc}
        \hline
        \textbf{} & \textbf{Morning Apartment } & \textbf{Scene 0000 } & \textbf{Scene 0012 } \\
        \hline
        Dimension      & 3.8×2.9×4.8 & 9.6×9.6×3.8 & 6.7×6.7×3.8\\
        Voxel Dim & 129×97×161 & 321×321×129 & 225×225×229  \\
        Runtime  & 36min & 73min & 59min \\
        Model Size      & 82MB & 535MB & 263MB \\F
        Num Params         & 21.5M &  140.4M & 69.1M\\
        \hline
    \end{tabular}
    \label{tab:parmetersAnalysis}
\end{table}
\begin{table}[!t]
    \centering
    \scriptsize
    \caption{The performance comparison of PreSem-Surf with the baseline model after removing different functional modules}
    \begin{tabular}{lcccccc}
        \hline
        \textbf{Model} & \textbf{C-L1 $\downarrow$} & \textbf{NC $\uparrow$} & \textbf{F-score $\uparrow$} & \textbf{IoU $\uparrow$} & \textbf{Acc $\downarrow$} & \textbf{Comp $\downarrow$} \\
        \hline
        GO-Surf      & \underline{0.0398} & \underline{0.9209} & 0.9059 & 0.5358 & \textbf{0.0155} & \textbf{0.0649} \\
        No-Semantic & 0.0409 & \textbf{0.9210} & 0.9062 & \underline{0.5370} & 0.0169 & \underline{0.0664} \\
        No-SG-MLP  & 0.0462 & 0.9152 & \underline{0.9071} & \underline{0.5176} & 0.0218 & 0.0708 \\
        \hline
        PreSem-Surf       & \textbf{0.0229} & 0.9193 & \textbf{0.9186} & \textbf{0.5629} & \underline{0.0165} & 0.0735 \\
        \hline
    \end{tabular}
    \label{tab:3}
\end{table}

\subsection{Reconstruction Quality}
As can be observed from the scene reconstruction test results, shown in Fig. \ref{fig:qualitative-comparison} and Table \ref{tab:performanceCompa}, GO-Surf achieved good smoothness in the reconstruction results, but it lacks a fine depiction of scene details, and the reconstruction effect appears somewhat bloated.
Fragmentation problems occur in areas such as the chair backs in the breakfast room and complete kitchen, the picture frames in the green room, the windows in the grey-white room and white room, and the cabinets in the morning apartment.
Neural-RGBD achieved good reconstruction smoothness with almost no fragmentation, but there are visible misalignment issues.
Moreover, Neural-RGBD is prone to misjudgment. For example, in the grey-white room scene's window part, when there is a dense absence of depth data, the reconstruction result is poor, or there is often a lack of or excessive reconstruction in the reconstruction of objects such as cups and table legs.
PreSem-Surf, on the other hand, has achieved a good balance between smoothness and detail depiction, showing good performance in both aspects across the seven scenarios.
\subsection{Quantitative Analysis}

We employs the metrics C-L1, F-score, IoU, NC, Acc, and Comp to evaluate the reconstruction effectiveness of different models as comprehensively as possible. As shown in Table \ref{tab:3}, PreSem-Surf achieved satisfactory results.

Specifically, first, PreSem-Surf significantly outperformed or matched the benchmark in all metrics, indicating that the model can effectively address the 3D reconstruction problem. Second, PreSem-Surf achieved the best performance in C-L1, F-score, and IoU, suggesting that the model's reconstruction results have high precision, can well reflect the actual situation of the scene, and have achieved a good balance between precision and recall. Third, PreSem-Surf's performance in NC, Acc, and Comp was marginally inferior to the best models, indicating that the model performs well in handling normal consistency and also has a good performance in the accuracy and completeness of point cloud reconstruction.

In summary, PreSem-Surf achieves high-precision reconstruction while also taking into account the smoothness and completeness of the reconstruction, demonstrating excellent comprehensive performance.

\begin{figure}[t]
    \centering
    \begin{tikzpicture}
        \centering
        % 使用tikz绘制一个带边框的节点，包裹整个minipage
        \node[xshift=0cm, yshift=0cm, draw = none, thick, inner sep=5pt] { % 设置框线厚度和内部间距
            \begin{minipage}{0.45\textwidth} % 控制图像宽度，0.45\textwidth即为45%的页面宽度
                \centering
                \begin{subfigure}[t]{0.3\textwidth} % 调整每个子图的宽度
                    \centering
                    \includegraphics[width=\textwidth]{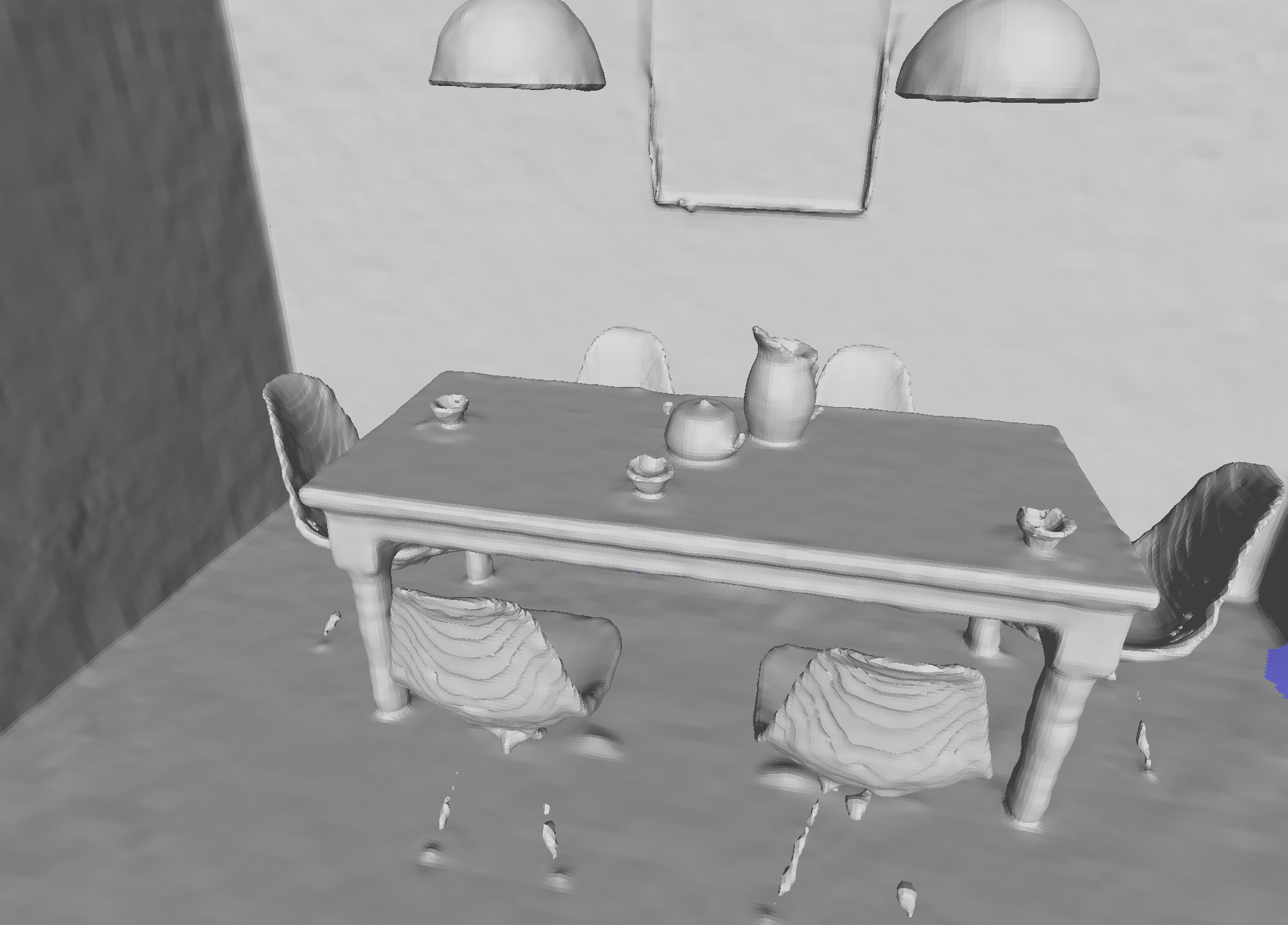}
                \end{subfigure}
                \begin{subfigure}[t]{0.3\textwidth}
                    \centering
                    \includegraphics[width=\textwidth]{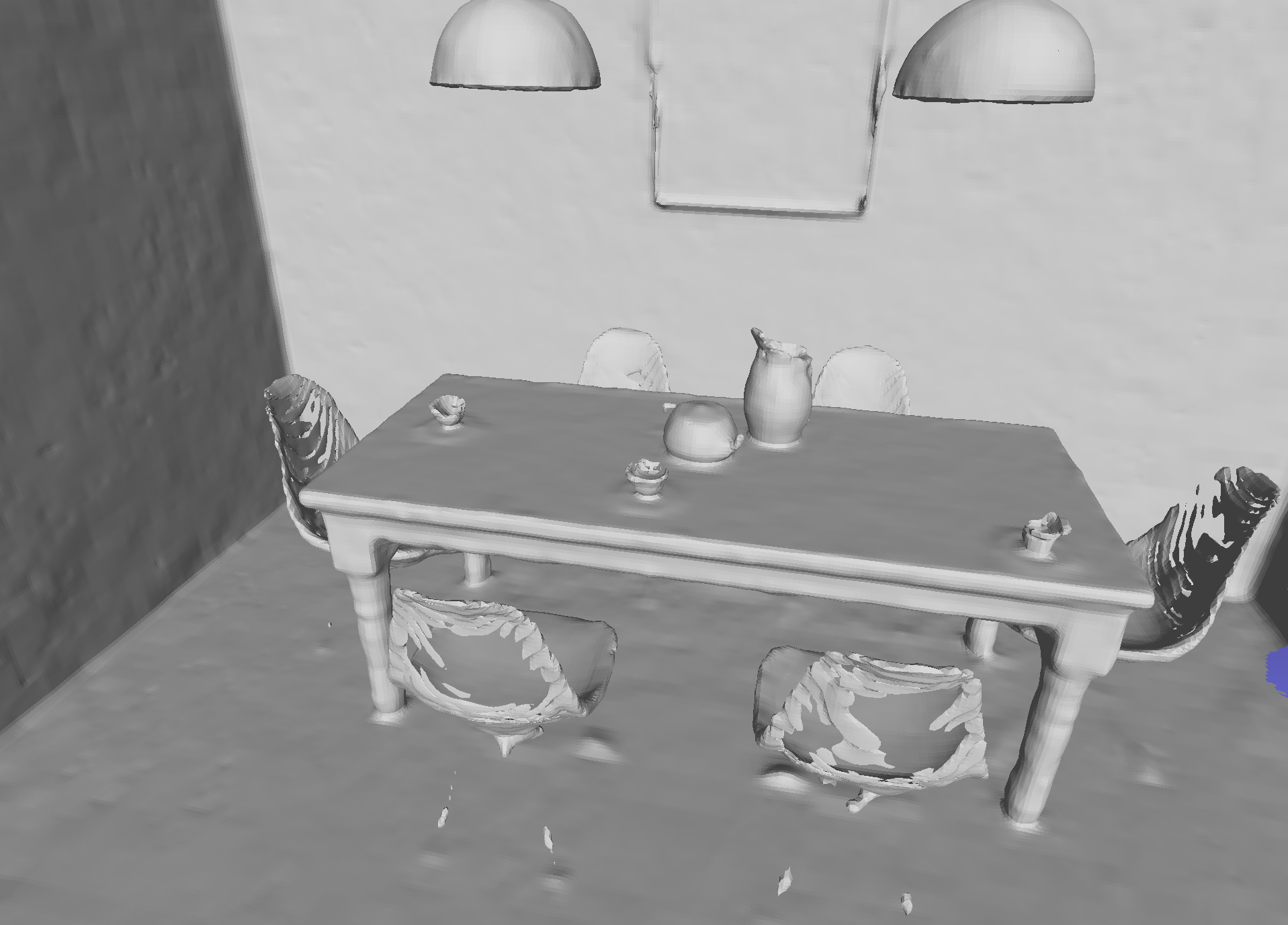}
                \end{subfigure}
                \begin{subfigure}[t]{0.3\textwidth}
                    \centering
                    \includegraphics[width=\textwidth]{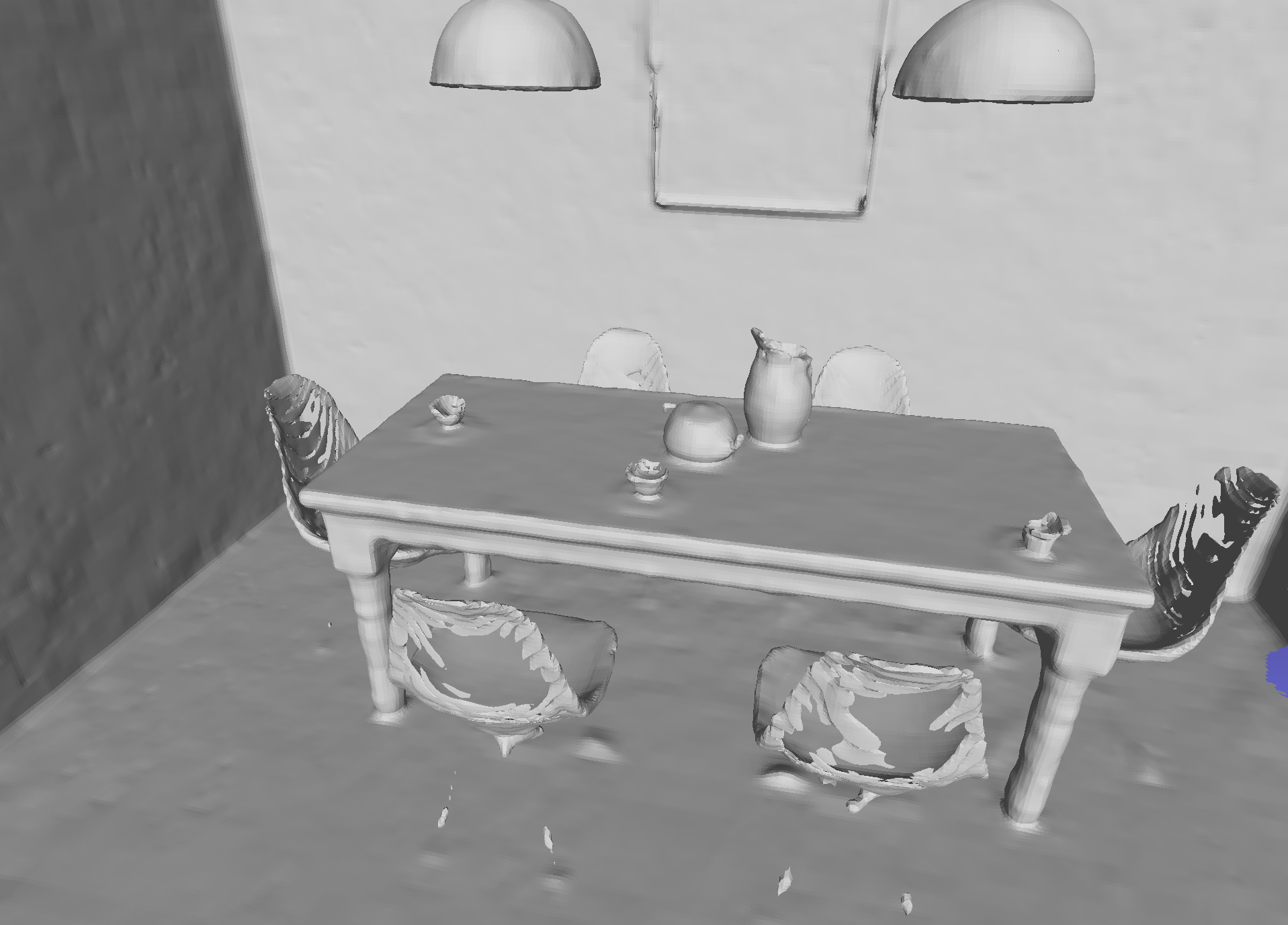}
                \end{subfigure}
                \\ % New row
                % Second row
                \begin{subfigure}[t]{0.3\textwidth}
                    \centering
                    \includegraphics[width=\textwidth]{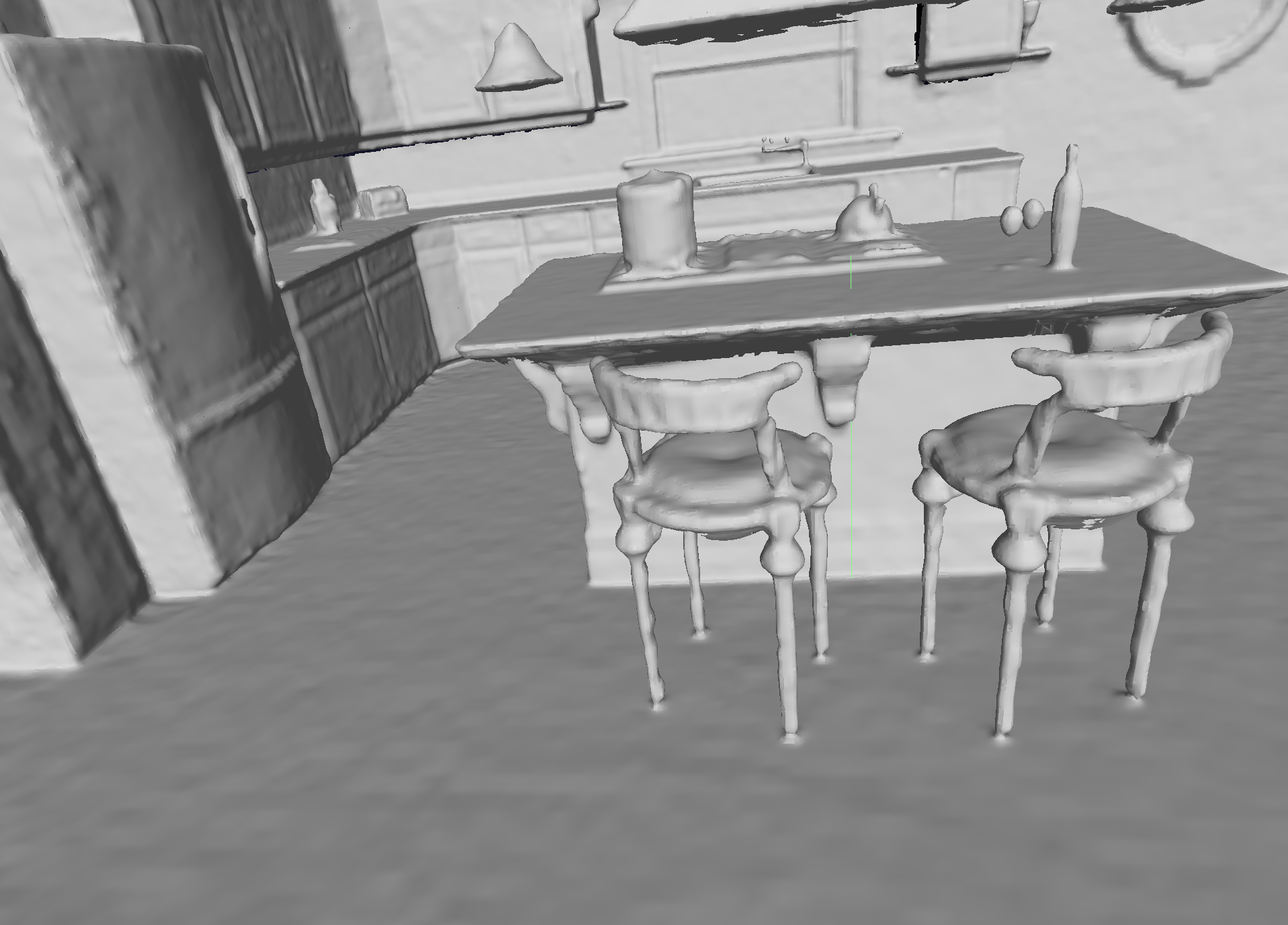}
                \end{subfigure}
                \begin{subfigure}[t]{0.3\textwidth}
                    \centering
                    \includegraphics[width=\textwidth]{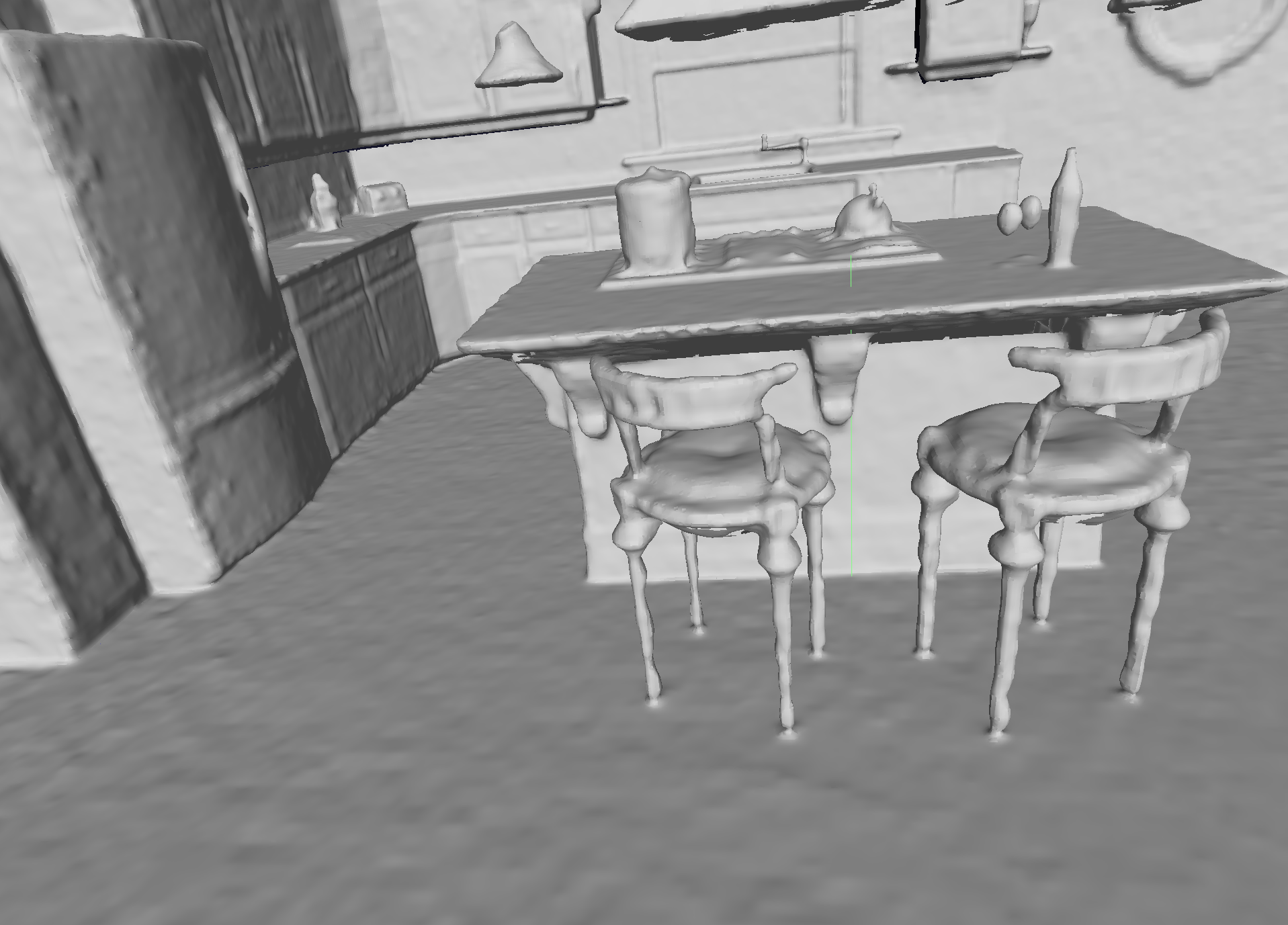}
                \end{subfigure}
                \begin{subfigure}[t]{0.3\textwidth}
                    \centering
                    \includegraphics[width=\textwidth]{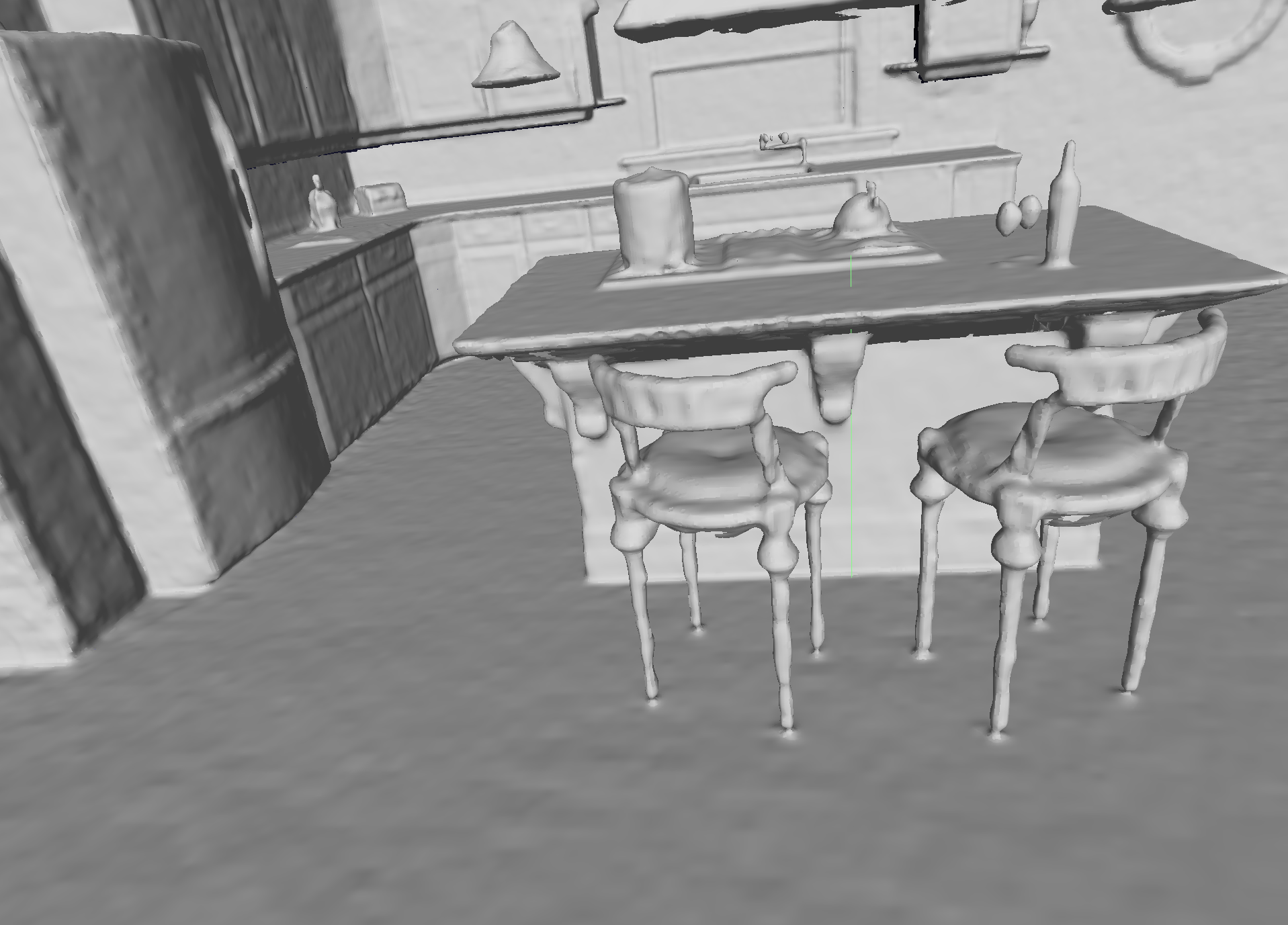}
                \end{subfigure}
                \\
                % Third row
                \begin{subfigure}[t]{0.3\textwidth}
                    \centering
                    \includegraphics[width=\textwidth]{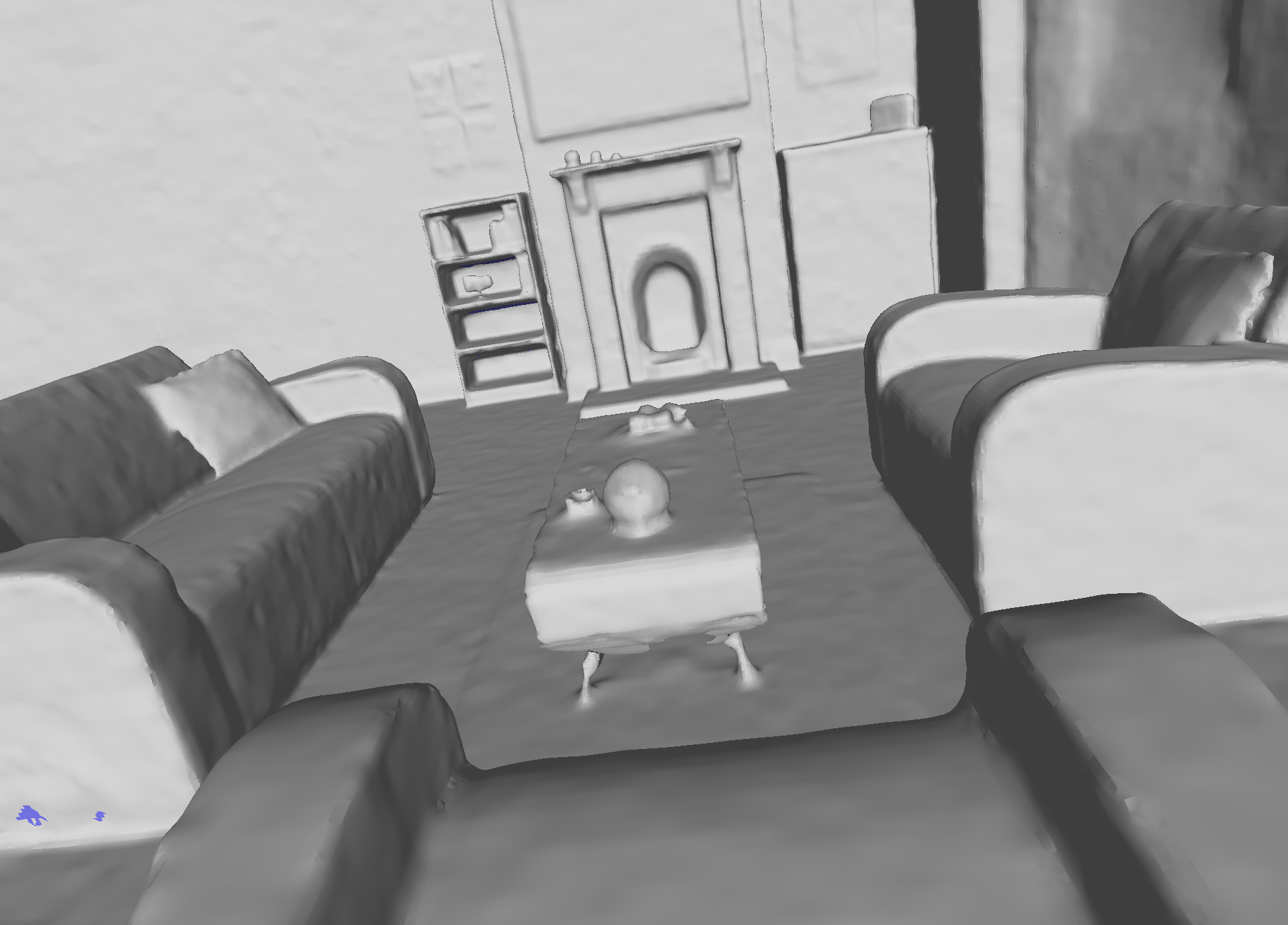}
                    \caption*{\scriptsize PreSem-Surf}
                \end{subfigure}
                \begin{subfigure}[t]{0.3\textwidth}
                    \centering
                    \includegraphics[width=\textwidth]{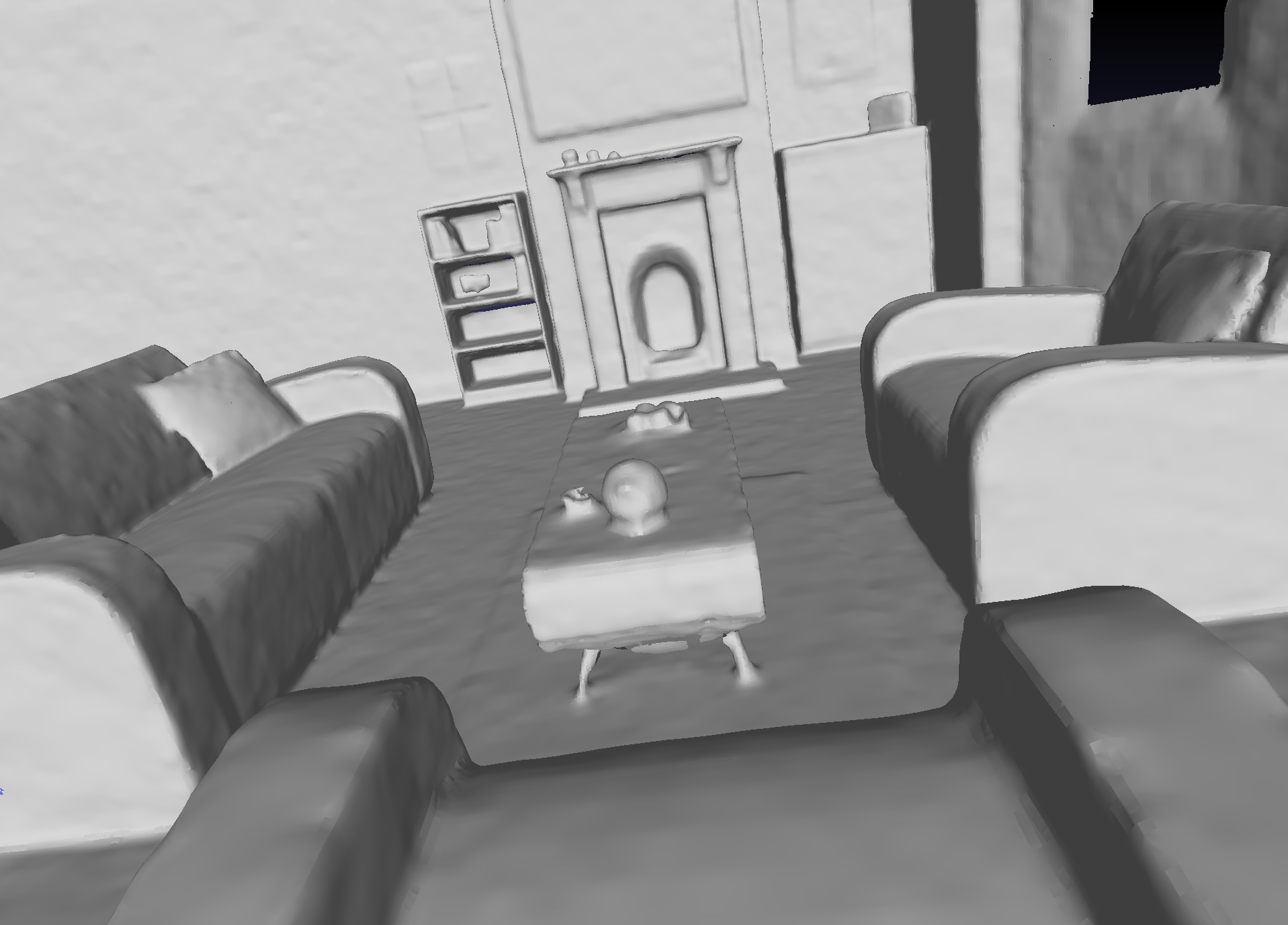}
                    \caption*{\scriptsize No-Semantics}
                \end{subfigure}
                \begin{subfigure}[t]{0.3\textwidth}
                    \centering
                    \includegraphics[width=\textwidth]{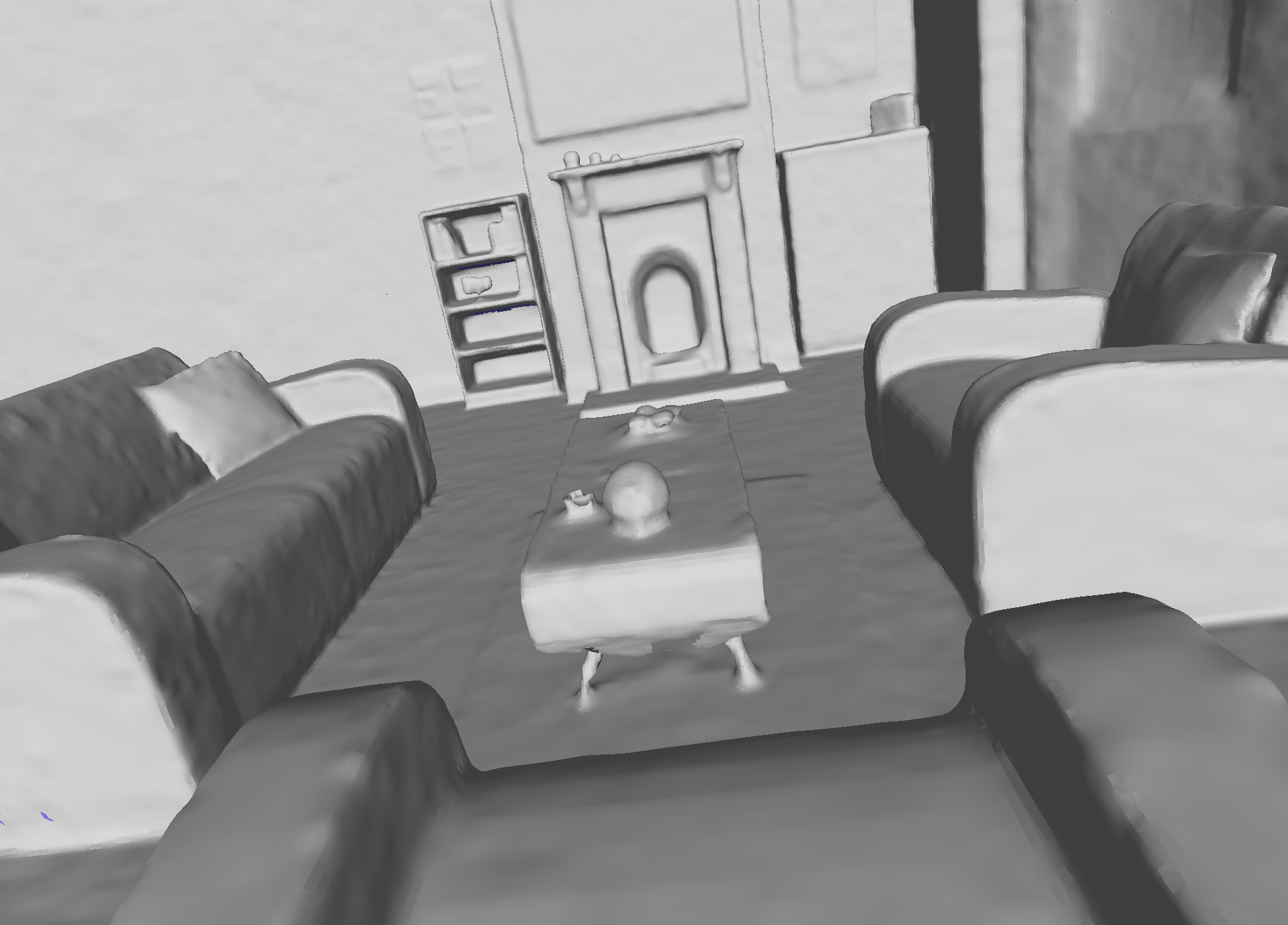}
                    \caption*{\scriptsize No-SG-MLP}
                \end{subfigure}
                \\
                \caption{The visualization performance of PreSem-Surf after removing different functional modules.}
            \end{minipage}
        };
    \end{tikzpicture}
\end{figure}
\subsection{parameters analysis}

We selected the "Morning Apartment" from the Synthetic Dataset. Additionally, we chose scenes 0 and 2 from ScanNet. This was done to calculate the performance metrics of PreSem-Surf in both simulated and real scenarios. As shown in Table \ref{tab:parmetersAnalysis}, our model performs well overall. However, its memory usage and time cost increase rapidly with the scale of the scene, which is a common drawback of voxel-based models. How to further optimize this is a direction for our future research.
\subsection{Ablation Studies}
We conducted ablation studies on our model across different scenarios to validate the impact of each module on the reconstruction effect and to substantiate the rationality and effectiveness of our design.
Impact of the SG-MLP module: As shown in Table \ref{tab:3}, after removing the SG-MLP module, the model's performance significantly declined on metrics such as C-L1 and IoU, indicating that the model lost some understanding of the overall scene structure. This proves to some extent that the SG-MLP can assist the model in grasping global scene information, which aligns with our initial design intent.
As demonstrated in Table \ref{tab:3}, the removal of Semantic Model resulted in a noticeable decrease in the model's performance on key metrics such as C-L1, F-score, and IoU. This indicates that the incorporation of Semantic Model plays a significant role in enhancing the overall reconstruction results.
However, we also observed improvements in certain metrics after the Semantic Model was removed, suggesting that while it improves the overall quality of reconstruction, it may introduce adverse effects in specific aspects. 

\section{Acknowledgment}
This work was supported in part by the Natural Science Foundation Project of Fujian Province 2023J01432; in part by Industry-Academy Cooperation Project under Grant 2024H6006; in part by the Collaborative Innovation Platform Project of Fuzhou City under Grant 2023-P-002; in part by the Key Technology Innovation Project for Focused Research and Industrialization in the Software Industry of Fujian Province; and in part by the Key Research and Industrialization Project of Technological Innovation in Fujian Province under Grant  2024XQ002.
\section{Conclusion}

We propose PreSem-Surf, an innovative, efficient method capable of reconstructing high-quality surfaces from RGB-D sequences. This method integrates RGB image information, depth data, and rich semantic information. It innovatively designs a Sampling-Guided Multi-Layer Perceptron for hierarchical sampling and rendering. Furthermore, through the PFPSMS, scene reconstruction is carried out from coarse to fine based on semantic information, allowing the model to achieve more precise and complete scene reconstruction while significantly reducing training time.

% \vspace{12pt}
% \color{red}
% IEEE conference templates contain guidance text for composing and formatting conference papers. Please ensure that all template text is removed from your conference paper prior to submission to the conference. Failure to remove the template text from your paper may result in your paper not being published.

\end{document}